\documentclass{article}

\usepackage{arxiv}

\usepackage[utf8]{inputenc}
\usepackage[T1]{fontenc}
\usepackage[english]{babel}
\usepackage{setspace}
\usepackage[section]{placeins}
\usepackage{xcolor}
\usepackage{breakcites}
\usepackage{hyphenat}
\usepackage{float}
\usepackage{graphicx}
\usepackage[labelfont=bf]{caption}
\usepackage{subcaption}
\usepackage{latexsym}
\usepackage{textcomp}
\usepackage{longtable}
\usepackage{tabulary}
\usepackage{booktabs,array,multirow}
\usepackage{threeparttable}
\usepackage{amsfonts,amsmath,amssymb}

\newcolumntype{P}[1]{>{\raggedright\arraybackslash}p{#1}}

\usepackage[round,authoryear]{natbib}
\setcitestyle{aysep={}, notesep={: }}
\usepackage[resetlabels]{multibib}
\newcites{appendix}{References}
\bibliographystyleappendix{asr}
\usepackage{endnotes}

\PassOptionsToPackage{hyphens}{url}
\usepackage[colorlinks = true,
            linkcolor = blue,
            urlcolor  = blue,
            citecolor = blue,
            anchorcolor = blue]{hyperref}
\usepackage{orcidlink}

\title{Forced Displacement of People Experiencing Homelessness:\\
Housing and Movement Outcomes after Encampment Clearances}

\renewcommand{\shorttitle}{Housing and Movement Outcomes after Encampment Clearances}
\renewcommand{\headeright}{A Preprint}
\renewcommand{\undertitle}{A Preprint}

\author{Brandon Morande \\
	University of Washington\\
	\And
	Kim Serry \\
	Evergreen Treatment Services REACH\\
	\And
	Amy Hagopian \\
	University of Washington\\
	\And
	Zack W. Almquist\,\orcidlink{0000-0002-1967-123X}\thanks{Corresponding author: \texttt{zalmquist@uw.edu}.} \\
	University of Washington
}
\date{}

\begin{document}

\maketitle

\begin{abstract}
\noindent The 2024 \textit{Grants Pass} decision newly emboldens US cities to manage unsheltered homelessness through forced displacement. Although literature demonstrates the harmful health and material impacts of this tactic, the longer-term results for housing outcomes and migration patterns remain less clear. In response, this study leverages longitudinal street outreach data to investigate where people move following encampment clearances. We specifically employ relational event models to predict the likelihood of various outcomes post-removal, such as relocating tracts, entering shelter, or obtaining housing. Results suggest that displaced residents do not travel far, yet clearances may still reduce visible homelessness by decreasing the size of camp communities. Furthermore, people appear unlikely to move indoors and instead face high risks of losing contact with service providers. These trends hold regardless of individual demographics, although people with mental health conditions demonstrate stronger attachments to their original sites. Evidence additionally indicates that neighborhood conditions could influence these migration behaviors. Such findings corroborate broader literature on place attachments, residential mobility, and invisibilization of poverty. This paper ultimately addresses the urgent need for a deeper understanding of how forced displacement impacts homelessness and pathways to housing.
\end{abstract}


\section*{Introduction}

Homelessness has reached unprecedented levels in the United States, growing increasingly salient across many urban landscapes. According to recent estimates, over 270,000 people slept unsheltered on a single night in January 2024, often occupying publicly visible locations, such as sidewalks, vehicles, and parks \citep{de_sousa_2024_2024}. Cities have historically relied on arrests, trespass ordinances, and other exclusionary policies to help conceal this crisis \citep{beckett_banished_2010, mitchell_homelessness_2011}. Yet amidst pressure from advocacy groups, municipalities have increasingly invested in non-police alternatives and harm reduction services to address the problem \citep{beckett_diversion_2023, collins_pandemic_2023}. Even when avoiding more punitive legal sanctions, cities nevertheless still actively displace encampments and vehicle residences -- often mobilizing law enforcement for such events \citep{herring_complaint-oriented_2019, goldfischer_encampments_2020, margier_compassionate_2023, pruss_punitive_2020}.

Proponents argue that these interventions disincentivize unhealthy behaviors and encourage people to accept assistance \citep{herring_pervasive_2020, robinson_no_2019}. Officials specifically justify encampment clearances as compassionate measures that prioritize public health, community safety, and inhabitant vulnerability, especially when coupled with outreach services \citep{city_of_seattle_one_2023, margier_compassionate_2023}. However, literature suggests that clearances often worsen peoples’ security, service use, support networks, and legal involvement. Individuals also frequently lose personal belongings and survival supplies during such events \citep{darrah-okike_it_2018, herring_complaint-oriented_2019, margier_compassionate_2023, robinson_no_2019}. These impacts seem particularly drastic for those with mental illness, substance use disorder, and other chronic conditions \citep{barocas_population-level_2023, chang_harms_2022, chiang_health_2022, goldshear_notice_2023, qi_health_2022}. Some scholarship even suggests that clearances remain inadequate at bringing people indoors \citep{giamarino_echoes_2024, hayes-chaffe_audit_2023, herring_pervasive_2020}. Such inefficacy lends credibility to counterarguments that spatial removal may partly reflect other pressures to manage urban aesthetics \citep{blau_visible_1992, collins_policing_2022, gibson_securing_2004, margier_compassionate_2023, peck_neoliberal_2009, smith_new_1996, speer_urban_2019}.

Despite little evidence of positive outcomes for displaced residents, cities nevertheless continue relying on encampment clearances. In light of the US Supreme Court recently upholding civil and criminal penalties for camping on public land \citep{noauthor_city_2024}, this study meets the urgent need to better understand this tactic's consequences. Although literature has extensively documented the immediate material, health, and social harms of such practices, few studies examine peoples' experiences and pathways following displacement. These gaps partially result from inadequate documentation of removal practices and tracking of affected individuals. Additionally, existing scholarship primarily relies on interviews, surveys, and other cross-sectional methods that remain susceptible to recall bias and may not readily account for intermediary events or mechanisms \citep{chang_harms_2022, robinson_no_2019, darrah-okike_it_2018, herring_complaint-oriented_2019}. In response to these limitations, the American Public Health Association \citeyearpar{apha_protecting_2023} has notably called for more comprehensive research ``to address the lack of evidence-based literature surrounding forcible displacement of encampments" (p. 10).

Addressing such gaps, this study leverages longitudinal street outreach data to investigate people's housing and movement patterns after clearances in Seattle, Washington. We use relational event models to compare the likelihood of various outcomes for displaced residents, such as entering shelter, obtaining housing, or migrating between areas. We find that individuals typically do not relocate far, corroborating literature on place attachments and geographically-constrained mobility in high-poverty settings. Nevertheless, clearances still appear to reduce visible homelessness by severing camp communities. Evidence further suggests that people face much higher risks of losing contact with service providers compared to moving indoors. These trends persist across demographic characteristics, though mental illness and substance use seem predictive of remaining tethered to origin tracts. Migration patterns may additionally reflect neighborhood conditions, with clients in denser mixed-use districts showing greater risks of shuffling between nearby areas. These findings ultimately extend beyond the harmful effects of spatial removal, encouraging communities to implement evidence-based approaches to increase housing equity.

\section*{Literature Review}

The displacement of homelessness represents a long-standing tactic to conceal poverty \citep{blau_visible_1992} comparable to other forms of residential segregation that have historically confined poor residents \citep{desmond_evicted_2016, harrington_other_1962, riis_how_1890}. Although the US Supreme Court invalidated laws that broadly criminalized vagrancy in 1972 (\citeauthor{noauthor_papachristou_1972}), cities began creating new civility codes in the 1980s to address rising homelessness. These codes granted police authority to arrest individuals for \textit{specific} behaviors, such as sitting on sidewalks \citep{beckett_banished_2010}. Some municipalities thereafter broadened these interventions. In Seattle, for example, park exclusion ordinances, trespass admonishments, and other policies allowed officials to ban people from wide swaths of the city for infractions related to poverty and behavioral health \citep{beckett_banished_2010}. Throughout the 2000s, municipalities across the country increasingly employed these laws to contain marginal populations in liminal areas (e.g., skid rows) \citep{gibson_securing_2004, stuart_down_2016}, as well as constrict spaces for unsheltered survival \citep{mitchell_homelessness_2011}.

Rather than simply incarcerate, these ordinances seek to shape the geographic and institutional circulation of unhoused residents. Instead of a punitive ``revancishm" against the poor \citep{smith_new_1996}, scholars have documented more ambiguous practices over the past decade \citep{deverteuil_post-revanchist_2019}. In Los Angeles, for example, Stuart \citeyearpar{stuart_down_2016} observes that law enforcement officers increasingly engage in ``therapeutic policing," utilizing threats of arrest to compel individuals into shelter or treatment rather than jail -- notably at the behest of non-profit providers. Herring and colleagues \citeyearpar{herring_complaint-oriented_2019,herring_pervasive_2020,herring_complaint-oriented_2021} similarly demonstrate that authorities in San Francisco respond to the majority of residential and business complaints about homelessness by informally pressuring people to relocate and, more recently, enter shelter. These interventions often include other non-police actors (e.g., sanitation teams) and may represent preferable responses for cities amidst rising public pressure and limited resources. Tellingly, 90\% of interviewees sleeping outside and 80\% of those in vehicles reported being recently displaced by such methods \citep{herring_pervasive_2020}.

Authorities may thus increasingly respond to homelessness by forcibly shuffling people around space without further legal punishment. Admittedly, cities across the US slowed or halted these practices during the COVID-19 pandemic and Black Lives Matter protests. Some jurisdictions even invested in more supportive services, such as low-barrier shelter, non-police crisis teams, and Housing First programs \citep{collins_pandemic_2023, colburn_hotels_2022}. Yet this reprieve may represent a temporary anomaly, given that many cities experienced post-pandemic demands for punitive policing as business districts reopened \citep{collins_pandemic_2023}. This resurgence has notably included the targeting of unauthorized encampments, with at least 65 cities criminalizing or removing these sites as of early 2022 \citep{cline_liberal_2022}. For our purposes, encampments consist of temporary structures or enclosed places not intended for long-term habitancy, which people occupy on a continuous basis \citep{cohen_understanding_2019,dunton_exploring_2020}. Seattle’s Finance and Administrative Services (FAS) Rule 17-01 \citeyearpar{FAS_2017} similarly identifies encampments as ``one or more tent[s], structure[s], or assembl[ies] of camping equipment or personal property” in active use.

Public authorities and media outlets often criticize encampments for safety, health, and environmental concerns \citep{bawarshi_media_2008,beach_discontent_2024,city_of_seattle_one_2023}. However, scholars note that officials may weaponize such discourse (alongside expanded services) to justify clearances \citep{hennigan_compassionate_2019, herring_complaint-oriented_2021, speer_its_2017}. In Portland, Oregon, for example, authorities mobilize street outreach workers with police to evict residents from highly visible camps \citep{margier_compassionate_2023}. Margier \citeyearpar{margier_compassionate_2023} attests that this ``compassionate invisibilization” (p. 193) allows the city to appease public complaints, while pacifying advocates opposed to harsher penalties. Across the west coast, shelter and housing providers even promote sweeps as a means to reduce perceived disorder, improve urban aesthetics, and gain support for their programs \citep{hennigan_compassionate_2019, herring_complaint-oriented_2021, speer_its_2017}. Clearances thus provide a fruitful arena for evaluating how rhetoric and practices of care supersede, complement, or belie those of punishment \citep{deverteuil_complexity_2009}. Indeed, the Seattle Mayor's Office notably refers to these tactics as ``resolutions" \citep{city_of_seattle_one_2023}, reflecting stated goals of facilitating access to services.

Per such claims, we would expect to observe beneficial outcomes for displaced residents. Yet literature both domestically and abroad suggests otherwise, particularly given the lack of alternative residence options for this population. Based on Point-In-Time counts, the US does not operate enough shelter beds to serve everyone experiencing homelessness \citep{junejo_no_2016}. Furthermore, the inaccessibility of shelter for many individuals (e.g., people with families, substance use disorder, criminal history) exacerbates this shortage. For those who cannot access such accommodations, encampments can offer more safety, privacy, and stability than sleeping elsewhere outside. In some cases, they can provide greater security, community, and autonomy than shelters \citep{junejo_no_2016, chang_harms_2022, herring_new_2014, wusinich_if_2019}. Apart from meeting survival needs, mutual support in camps may even bolster emotional buffers against stigmatization and discrimination from the broader public \citep{boucher_they_2022}. People residing in well-managed and legally-sanctioned sites can also focus on long-term plans, rather than navigating daily pressures of finding places to sleep safely and store belongings \citep{junejo_no_2016, herring_new_2014}.

In addition to not offering preferable housing arrangements, cities can cause substantial harm through actions associated with displacement. Clearing unauthorized encampments typically involves evicting inhabitants, removing personal property, and disbanding temporary structures \citep{apha_protecting_2023,chang_harms_2022}. Affected residents may not receive advanced notice or shelter offers \citep{apha_protecting_2023,goldfischer_encampments_2020}, and they often lose crucial belongings, including survival supplies, identification, appointment cards, and medications \citep{darrah-okike_it_2018, chang_harms_2022, herring_pervasive_2020}. In Honolulu, Hawaii, for example, 51\% of interviewees reported loss of IDs during removal events, with most unable to retrieve such documents due to the distance of storage sites, associated fees, and lack of instructions for retrieval \citep{darrah-okike_it_2018}. Refusing to relocate may also result in arrest, criminal summons, and civil citations \citep{chang_harms_2022, robinson_no_2019}, as well as parking-related tickets and impoundment for vehicle residents \citep{pruss_punitive_2020}. These material hardships and legal obligations can subsequently pose obstacles to work, education, and housing \citep{herring_complaint-oriented_2019}.

Clearances can also worsen physical well-being by decreasing safety and disrupting access to care. The lack of stable housing in the US already exacerbates risks of morbidity and mortality. Tellingly, the mean age of death among people experiencing homelessness in Seattle was 48.9 years old between 2009 and 2019, compared to an average age of death of 80.3 years across Washington \citep{tanous_locations_2024}. Researchers suggest that involuntary displacement might contribute to this discrepancy \citep{chang_harms_2022}. For example, camping bans and relocation appear correlated with greater odds of violence, life-threatening health outcomes, and high-risk behaviors \citep{chiang_health_2022}. Displacement may also increase the prevalence of sexual assault and rape for females, as well as feelings of insecurity for gender-variant people \citep{herring_pervasive_2020}. Furthermore, forced relocation can augment acute complications, create obstacles to attending out-patient care, and interfere with disease management due to loss of medical supplies \citep{goldshear_notice_2023, qi_health_2022}. Lastly, people report that camping bans cause them to sleep less, as they spend more time moving locations to avoid police \citep{robinson_no_2019}.

Additionally, encampment removals can negatively impact mental and behavioral health, with implications for long-term stability. Displaced residents often describe feeling surveilled and dehumanized during these events, as well as dejected due to worries about future clearances \citep{darrah-okike_it_2018, goldshear_notice_2023}. These interventions can also reduce trust within unhoused communities and increase interpersonal conflict when people relocate to unfamiliar camps \citep{chang_harms_2022,goldshear_notice_2023}. With respect to substance use, literature further suggests that removal disrupts use of harm reduction services, addiction counseling, and medication-assisted treatment \citep{apha_protecting_2023, chiang_health_2022}. Displacement additionally exacerbates risks of infections, hospitalizations, and overdose, as people lose clean supplies, cope via substances, and relocate to isolated locations \citep{apha_protecting_2023, chiang_health_2022, qi_health_2022, barocas_population-level_2023}. Simulation models even project that repetitive moves could contribute up to 24.4\% in additional deaths over a 10-year period among people experiencing homelessness who inject drugs, with a point estimate of 19\% for Seattle \citep{barocas_population-level_2023}.

Given these impacts, clearances have faced much legal scrutiny for violating constitutional rights regarding unreasonable seizures, cruel and unusual punishments, and unequal protections \citep{noauthor_lavan_2012, noauthor_ellis_2016,noauthor_coalition_2022}. Locally, the King County Superior Court judged that the City of Seattle's overly broad definition of ``obstruction" (which permits clearances without notice) rendered some removals unconstitutional \citep{noauthor_bobby_nodate}. However, the US Supreme Court recently determined that enforcement of public camping bans does not violate Eighth Amendment protections against cruel and unusual punishment \citep{noauthor_city_2024}. Cities can legally arrest and cite people for sleeping unsheltered, even in the absence of safe sleeping alternatives. As a result, some lower courts have reversed previous protections, and many states and cities have passed legislation to further criminalize homelessness (ACLU \citeyear{american_civil_liberties_union_one_2025}).

Considering the harmful effects of spatial removal, we therefore ask whether encampment clearances can truly facilitate pathways out of homelessness? Studies that quantify outcomes remain especially scarce and rarely track residents beyond displacement. Nonetheless, limited evidence indicates that people overwhelmingly remain outside. In Denver, 40\% of survey respondents reported increasing their efforts to use shelters following a camping ban \citep{robinson_no_2019}. However, most of these respondents experienced difficulty accessing a bed, with 73\% denied entry due to lack of capacity \citep{robinson_no_2019}. In San Francisco, only 9\% of survey respondents reported transitioning indoors following their most recent move-along order \citep{herring_pervasive_2020}. The New York City Comptroller similarly found that between March and November 2022, just 90 (3.9\%) of 2,308 people forcibly removed from encampments remained in a shelter for more than a day, with solely three people obtaining permanent housing \citep{hayes-chaffe_audit_2023}.  Despite a multimillion budget for outreach workers, clearances in Portland similarly moved less than 15\% of displaced residents into shelter in a given month, while less than 5\% received housing referrals. Lastly, in Los Angeles' Echo Park Lake clearance from March 2021, only 17 (9.3\%) of 183 evicted residents entered housing after a year \citep{giamarino_echoes_2024, roy_displacement_2022}.

Where people migrate following displacement, however, appears less clear. In San Francisco, Herring et al. \citeyearpar{herring_pervasive_2020} find that most survey respondents (64\%) who received move-along orders stayed nearby or quickly returned to the same location, while only 21\% relocated to a different neighborhood. Instead of unidirectional movement into certain areas, the authors observe an even churning between districts, with the overall number of people in a space remaining relatively constant. In Honolulu, interviewees also often report returning to the same sites after displacement, where they feel safer due to support networks \citep{darrah-okike_it_2018}. The aforementioned New York City audit \citep{hayes-chaffe_audit_2023} similarly observed that approximately a third of locations cleared in 2022 had inhabitants in April 2023 -- though the report does not determine if the occupants remained the same. Qualitative research in Seattle likewise demonstrates that complying with trespass and banishment orders can feel impossible for individuals due to material and emotional attachments to certain areas \citep{beckett_banished_2010}. This literature further aligns with scholarship on the distance dependence of residential mobility, whereby most migration remains spatially constrained, especially if surrounding areas do not offer significant advantages \citep{clark_housing_1982, crowder_neighborhood_2011, lee_theory_1966, krysan_cycle_2017}.

Nevertheless, some studies suggest that encampment removals may still force people into environments further away from vital resources. In Denver, 66\% of survey respondents reported avoiding downtown following a camping ban, instead choosing more hidden spaces to sleep \citep{robinson_no_2019}. Chang et al. \citeyearpar{chang_harms_2022} similarly find that interviewees in Santa Clara County, CA, move to isolated and hazardous areas, such as along train tracks and freeways. In addition to restricting access to electricity, water, and bathrooms, these spaces render outreach from service providers increasingly difficult \citep{chang_harms_2022}. Such movement patterns may reflect the pervasive nature of clearances, whereby individuals repeatedly experience displacement \citep{chiang_health_2022, darrah-okike_it_2018} and removals occur frequently at the same sites \citep{goldshear_notice_2023}. Authorities may even install ``hostile architecture" (such as fencing) at these locations in order to deter future occupancy \citep{kim_jr_seattle_2020}. Given higher costs of remaining in such situations, displaced residents may relocate longer distances, even if this severs access to certain amenities.

Although valuably capturing unhoused experiences, these studies nevertheless display several limitations. Most rely on surveys or interviews that ask respondents to reflect upon past experiences \citep{chang_harms_2022,robinson_no_2019, herring_complaint-oriented_2019, darrah-okike_it_2018}. These methods could feasibly result in recall bias, especially in dynamic contexts of homelessness, where it may prove difficult to remember exact dates, locations, and timelines of events. Furthermore, other studies either employ cross-sectional population-based approaches that do not track specific clients \citep{hayes-chaffe_audit_2023} or case studies of single events \citep{giamarino_echoes_2024,roy_displacement_2022}. Expanding upon this literature, we leverage longitudinal data to more systematically assess people's trajectories following encampment clearances. We specifically evaluate whether displaced residents appear likely to transition indoors -- as proponents attest -- or instead lose service connections, remain unsheltered, and/or migrate elsewhere.

\section*{Study Site}

Seattle offers a compelling site to explore migration and housing patterns following displacement, offering trends that likely hold for other cities. King County, WA, recently recorded the fourth highest homelessness count among major metropolitan areas, with over 16,000 people unhoused on a given night (HUD \citeyear{hud_pit_2007-2024}).\endnote{This county-city comparison has received criticism, given that rates of homelessness often prove greater in urban relative to suburban areas \citep{colburn_homelessness_2022}. County numbers may thus underestimate the extent of this issue in Seattle compared to other cities (e.g., New York, Los Angeles).} Furthermore, 60\% of these individuals slept unsheltered \citep{kchra_2024_2024}, suggesting an especially visible population similar to numerous west coast cities. Like in California and Oregon \citep{herring_new_2014, margier_compassionate_2023}, this crisis has garnered considerable attention over the past two decades, with encampment clearances representing a key component of municipal policies. In 2005, King County's 10-year plan to end homelessness notably established guidelines for offering notice, services, and storage when removing camps \citep{cehkc_committee_to_end_homelessness_in_king_county_roof_2005}.\endnote{These guidelines ironically followed a decree that sanctioned some regulated tent cities as an emergency response \citep{cache_citizens_2004}. The City Council also passed legislation in 2011, legalizing encampments on church property. In 2015, the municipality further allowed temporary sites elsewhere.} A decade later, Mayor Murray declared a civil emergency against homelessness, dedicating over \$2 million to clearing sites \citep{murray_mayoral_2015}.  By 2017, the city created the Navigation Team (NT) to formally coordinate interventions between police, outreach staff, and sanitation workers.

During this period, Seattle additionally codified the Finance and Administrative Services (FAS) Rule 17-01 \citeyearpar{FAS_2017}, which permits clearances without 72-hour advanced warning if sites (1) block city properties or rights-of-way, (2) allegedly present risks of serious injury or death, or (3) were previously cleared and identified as consistent problems. According to public records, the city has provided advanced notice to increasingly fewer encampments, which seems to coincide with marked rises in interventions. FAS Rule 17-01 also established uniform procedures for removing sites from property managed by numerous city agencies, including Seattle Parks and Recreation, Public Utilities, and the Department of Transportation (SDOT). The related Multi-Departmental Administrative Rules (MDAR) 17-01 \citeyearpar{mdar_2017} prohibits specific uses on these properties, such as public access outside of operating hours, erecting unauthorized structures, and overnight camping -- except on designated sites (for parks) and non-specified areas (for SDOT). These policies effectively produce a \textit{de facto} municipal ban on encampments, functionally similar to \textit{de jure} restrictions enacted across the country (NLCHP \citeyear{nlchp_tent_2017}).

Clearances thereafter reached unprecedented levels, with the NT’s budget exceeding \$8 million by 2020. Eventually, the COVID-19 pandemic granted temporary relief from displacement, with the city pausing nearly all removals and later disbanding the NT.\endnote{Records obtained from the Mayor's Office through the Freedom of Information Act (FOIA) document approximately 90 removals from April 2020 through 2021, representing a substantial decrease from prior years (i.e., over 900 in 2019).} Yet this reprieve ended in 2022 when Mayor Bruce Harrell's administration created the Unified Care Team (UCT) to ``resolve encampments" \citep{city_of_seattle_one_2023}. Per public records, this multi-departmental coalition cleared around 750 sites in 2022 and 2100 in 2023. The UCT extended 1228 shelter offers to camp inhabitants in the final quarter of 2023 alone, of which 462 were accepted and led to referrals \citep{city_of_seattle_one_2023}. However, referrals do not necessarily result in placement or long stays \citep{herring_complaint-oriented_2021, stuart_down_2016}, and these figures fail to specify the total count of displaced residents. We nevertheless highlight these numbers to capture the growing scale of spatial removal, as well as the need for robust evaluation of outcomes. Fortunately, the increased formalization and documentation of clearances in Seattle provides researchers with rich data to more systematically study this issue, with potential implications for similar contexts across the US.

\section*{Data}

This paper focuses on the twelve largest clearances conducted by the City of Seattle between 2016 and 2018.\endnote{Given that our methods require larger displaced populations, we do not assess clearances under Harrell (2022 to 2024). Encampment communities remained much smaller during this period compared to prior years.} Following demands for data transparency and evidence-informed practices, the Human Services Department began publishing journals on city-directed inspections, cleanings, and removals of unauthorized encampments.\endnote{Readers can access the encampment site journals at \url{https://www.seattle.gov/human-services/reports-and-data/addressing-homelessness/encampments}. Based on public records obtained via FOIA, these documents do not cover all removal events, despite field coordinators being directed to complete them prior to any advanced notice or obstruction/hazard removals. Despite our requests, city officials did not clarify the discrepancies in the number of clearances, which could help determine potential data biases. We nevertheless use the journals given that they offer the most comprehensive details regarding geographic boundaries and the number of structures.} After downloading all documents from March 14, 2017 through April 24, 2024, we compiled a dataset of inspection and clean-up dates, site names, addresses, and numbers of structures. These counts include tents, bedrolls, vehicles, and additional undefined materials. Structures serve as a proxy for the population of encampments, given that residents were not always present during cleanings. The journals documented 653 clearances during this period,\endnote{The journals also listed 148 litter-pick services that did not result in eviction. Scholars have documented similar interventions elsewhere. For example, Los Angeles County has conducted ``spot cleanings," disposing of waste without removing tents or other belongings \citep{goldshear_notice_2023}.} from which we selected events with the greatest number of structures at inspection. We also included the notorious ``Jungle” clearance from 2016, which displaced hundreds of residents. The area witnessed a fatal shooting in January of that year, which prompted a final removal in October \citep{groover_as_2016}.

We then linked Evergreen Treatment Services REACH clients to these encampment clearances. Since 1996, REACH has connected individuals experiencing homelessness to shelter, medical care, and other resources in King County. The organization specifically operates an outreach program, in which staff regularly seek out and engage unsheltered individuals on the streets across the region. Staff document these encounters, including their times and locations, as well as client details. For this project, we drew removal boundaries for each clearance based on city journals, then combined adjacent events that occurred within a span of approximately one week. We thereafter isolated campsites within these boundaries that REACH visited up to 90 days prior to the final clearance date. This resulted in some overlap for removals occurring in close spatial and temporal proximity. Our team thereafter identified encampments bordering these boundaries and reviewed visit notes to clarify whether the clearances also covered such locations. Given sparse outreach data in 2019, we ultimately chose removal events from 2016 through 2018.

\begin{figure}[t]
    \centering
    \includegraphics[scale=0.65]{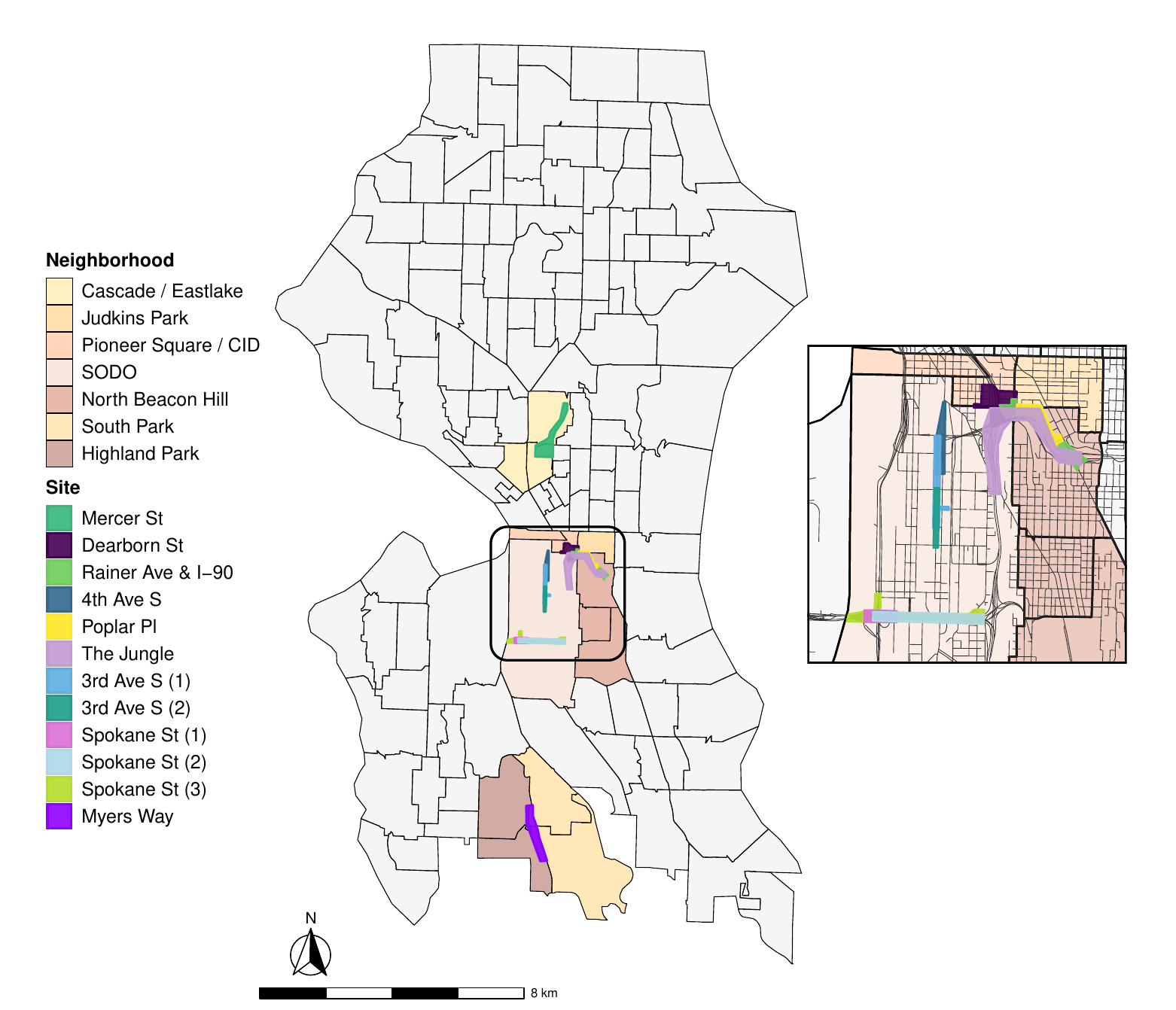}
    \caption{\textit{Clearance sites and associated neighborhoods.The neighborhoods represent pre-2020 Community Reporting Area (CRA) boundaries. The City of Seattle developed these spatial units as a standard community-oriented geography, which also aligns with Census tracts to facilitate data linkage \citep{city_of_seattle_cra_2024}.} Black lines on the large map represent Seattle's 2010 Census tract boundaries, while the zoomed inset also displays streets. The legends list neighborhoods and removals in descending order based on their northernmost points.}
    \label{fig:ref_map}
\end{figure}

Next, we queried unsheltered clients located at the sites within the 90-day windows, compiling data on their subsequent encounters with REACH and service providers across Washington connected to the Homeless Management Information System (HMIS).\endnote{Client data comes from an internal REACH database and broader Homeless Management Information System. The latter stores data on all homelessness services provided in King County, as well as other Continuums of Care (CoCs) in Washington State. As such, the final dataset should theoretically capture engagements with other providers both within and outside REACH's catchment area - although this depends on the privacy agreements between the specific client, entity, and CoC.} The dataset includes a client identifier, date, and location for each engagement, as well as individual demographics, health diagnoses, issues addressed, referrals made, living situations (e.g., streets), and death dates. For observations lacking coordinate information, we first isolated street and encampment engagements, since we could not assume unsheltered status from encounters at provider offices or institutional settings. We then standardized addresses due to non-uniform entry (e.g., of street suffixes) and used the \textit{tidygeocoder} package \citep{cambon_tidygeocoder_2021} in \textit{R} (v. 4.4.3; \citealp{R}) to supply the remaining coordinates.\endnote{We used the Coordinate Reference System (CRS) projection 4269 for \textit{tidygeocoder}, thereafter transforming coordinates into a 4326 projection based on REACH's format. For analysis, however, we utilized the NAD83 (National Spatial Reference System 2011) Washington North (EPSG: 6596) metric projection \citep{us_national_geodetic_survey_nad83_2022}.} With respect to client covariates, some attribute values consisted of strings of diagnoses, issues, and referral codes. We aggregated each phrase into categories (e.g., physical health) and created dummy variables. Lastly, we constructed a dataframe of clients' living situations and locations each day from (at most) three months before displacement up to a year after. Our analysis assumed inertia in coordinate positions between REACH encounters, as well as accounted for attrition (e.g., via death or lost contact). We also treated individuals who appear at multiple clearances as distinct data points. The final sample ultimately includes 468 unique client-clearance combinations across twelve events (Table \ref{tab:removal_sites}, Figure \ref{fig:ref_map}).

\begin{table}[t]
  \caption{\textit{Mass clearance events in Seattle, with dates and location information.}}
  \label{tab:removal_sites}
  \centering
  \begin{threeparttable}
      \begin{tabular}{@{\extracolsep{5pt}}lccc} 
\hline \hline \\[-1.8ex] 
Site Area & Removal End Date & Clients \\ 
\hline \\[-1.8ex] 
The Jungle & 10/12/16 & 78 \\
4th Ave S & 3/29/17 & 31 \\
Spokane St (1) & 4/14/17 & 22 \\
Rainier Ave \& I-90 & 5/25/17 & 76 \\
Spokane St (2) & 7/13/17 & 58 \\
Spokane St (3) & 9/15/17 & 58 \\
Mercer St & 5/10/18 & 19 \\
3rd Ave S (1) & 6/13/18 & 26 \\
Poplar Pl & 6/15/18 & 28 \\
Dearborn St & 6/27/18 & 33 \\
3rd Ave S (2) & 9/6/18 & 12 \\
Myers Way & 9/27/18 & 27 \\
\hline
\\[-1.8ex] 
& & n = 468
\end{tabular}
  \end{threeparttable}
\end{table}

\section*{Methods}

\begin{figure}[t]
\centering
\includegraphics[scale=0.55]{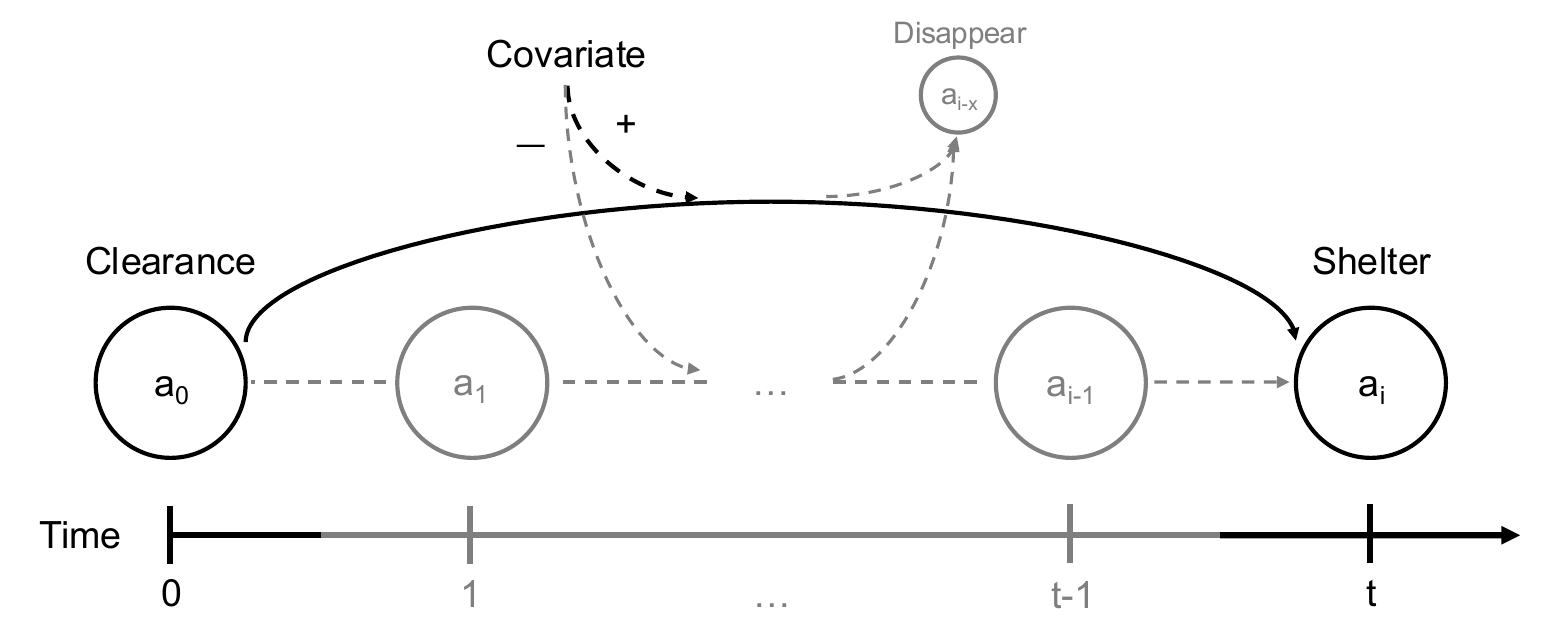}
\caption{\textit{Egocentric relational event model.} We predict the likelihood of an event ($a_i$, e.g., entering shelter) occurring at a given time ($t$) following a clearance ($a_0$). A \textit{potential} sequence of intervening events ($a_1$ to $a_{i-1}$) may mediate the risk of $a_i$. Individual and environmental covariates (e.g., gender) could also moderate this pathway. A client can exit the system by ``disappearing" or losing contact with service providers. Although not shown here, the model assesses the relative likelihoods of various possible $a_i$ events (e.g., moving tracts, obtaining housing).}
\label{fig:rem_concept_model}
\end{figure}

Our paper assesses the likelihoods of migration and housing outcomes in the year following displacement. This window grants outreach staff substantial time to locate clients, while acknowledging that eviction may cause prolonged instability. We specifically employ Butt's \citeyearpar{butts_relational_2008} relational event framework, which leverages time-ordered data to infer the risk of certain phenomena following an exogenous event, after accounting for complex dependencies (see Appendix for model notation). Although often used to examine group and dyadic interaction processes \citep{pilny_illustration_2016}, relational event modeling (REM) can also apply to individual actions \citep{marcum_constructing_2015}. This approach aptly suits relocation patterns in the context of unsheltered homelessness, where actors must respond quickly while confronting social instability, cognitive difficulties, information limitations, and other environmental constraints. We predict the risk of clients losing contact with service providers, moving between Census tracts (2010 boundaries), entering shelter, and/or obtaining housing. Additionally, we compare the likelihoods of event sequences, such as moving indoors after migrating to another tract. Finally, we isolate the effects of individual and environmental covariates on these patterns (Figure \ref{fig:rem_concept_model}).

With respect to model interpretation, REM predicts event hazards using the log-linear form. By exponentiating coefficients, we obtain a hazard ratio, interpreted as the relative risk of an event occurring compared to the base category. For covariates, these hazard ratios signify the multiplicative effect associated with a unit increase in the predictor variable. More positive hazards suggest higher likelihoods of occurrence \citep{marcum_constructing_2015}. We conduct these analyses in \textit{R} (v. 4.4.3; \citealp{R}) using the \textit{relevent} package \citep{relevent}. All models employ a Bayesian estimator with weakly informative priors ($\mu$ = 0, $\sigma$ = 100, $\nu$ = 4) that do not favor positive or negative log hazards. These parameters also permit extreme values without penalization, yet acknowledge the lower likelihood of outlying observations by specifying relatively heavy tails for the Student's t-distribution. Even with these weaker priors, the Bayesian approach provides more stable coefficients and helps avoid overfitting due to small sample sizes for certain event combinations. In order to isolate trajectories preempted by displacement, the final REM sample only includes clients confirmed as unsheltered at removal (n = 395), since clients may disappear or move indoors prior to a clearance notice for unrelated reasons. The only removal with a prolonged advanced warning -- the Jungle -- had solely two REACH clients (2.6\% of that subsample) moving indoors before the final clearance day (both at least a week prior). As such, we do not believe this sampling criteria significantly biases our results.

Our study further examines the extent to which spatial removal displaces unsheltered residents and reduces the visibility of homelessness. Although researchers have assessed the number and stability of tent locations in Seattle \citep{snedker_tent_2025}, none have systematically tracked those living in these communities. Addressing this gap, we evaluate the Euclidean distances traveled by individuals throughout the following year. These analyses account for attrition from the unsheltered sample over time (e.g., via entering housing) and assume inertia between outreach encounters, although individuals could theoretically move sleeping locations. We then assess changes in the size of encampment populations at various periods preceding and following clearance events. Describing the number and proportions of residents at sites can help capture how spatial removal diffuses homelessness and camp communities. Though clients could settle into a space with non-sampled residents, we can still gauge whether they reside with others from their initial site.

Throughout our analyses, we employ both distance- and tract-based measures of migration given their distinct implications. For example, tract boundaries often follow features that could potentially constrain mobility, such as highways, railroads, and water bodies \citep{us_census_bureau_glossary_2022}. Tracts are also designed to represent demographically homogeneous units and have conventionally been employed to study residential mobility and neighborhood conditions \citep{crowder_neighborhood_2011, lee_beyond_2008}. Relocating to another tract may thus entail traversing geographic and social boundaries that meaningfully impact peoples' daily experiences and resources. In contrast, egocentric distance buffers assume more freedom of movement across space \citep{lee_beyond_2008}, which might prove faulty in Seattle given its hilly and water-bound terrain, as well as discontinuities in street networks. Yet tract sizes can vary substantially within cities and, in some instances, migrating between them could simply entail crossing a residential street (see \citealt{lee_beyond_2008} for critiques). Using both measures allows us to navigate these competing concerns, and our findings fortunately converge across either approach.

\section*{Results}

Our findings overwhelmingly suggest that unsheltered residents are unlikely to move indoors after a clearance, compared to other outcomes. Figure \ref{fig:rem_simp} shows the hazard ratios, or relative risks, of certain events following removal (see Appendix for all model results). Note that the ratios never surpass 1, given that they are based on the reference category of displacement. Since every client experiences removal, all other events are less likely to occur. Yet we can still directly compare the risks of the other options. For example, clients are significantly more likely to ``disappear" or lose contact with Washington HMIS service providers compared to any other outcome (p $<$ 0.05). Indeed, the risk of disappearing is an estimated 14 times greater than entering shelter (95\% CI 7.3 to 27.2) and 20 times that of obtaining housing (95\% CI 9.6 to 45.3). Furthermore, we observe higher likelihoods of any unsheltered event (i.e., moved, remained in, or returned to tracts) compared to transitioning indoors. For example, we estimate a nearly 7 times higher risk of relocating to a non-origin tract versus a shelter (95\% CI 3.5 to 14.1). However, we cannot confidently distinguish between the hazards of moving tracts or remaining in an initial tract after displacement, as indicated by the overlapping credible intervals. The difference between entering shelter versus housing also does not appear statistically significant.

\begin{figure}[t]
\centering
\includegraphics[scale=0.7]{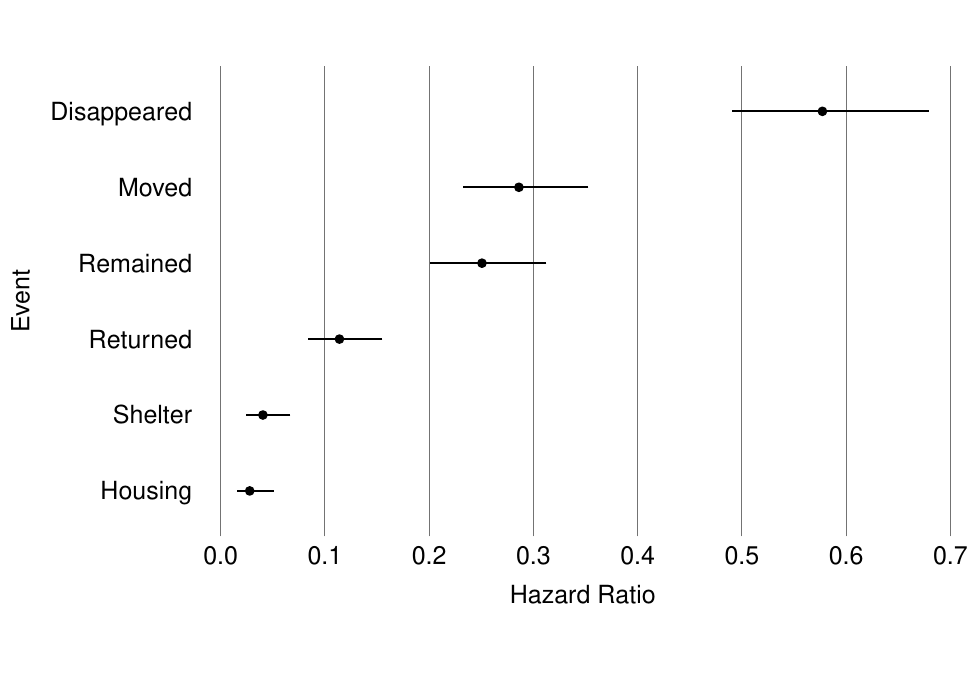}
\caption{\textit{Relative risks of certain events within a year after displacement.} Posterior mode estimates and 95\% credible intervals. Higher ratios
suggest greater likelihoods of occurrence.}
\label{fig:rem_simp}
\end{figure}

\begin{figure}[t]
\centering
\includegraphics[scale=0.7]{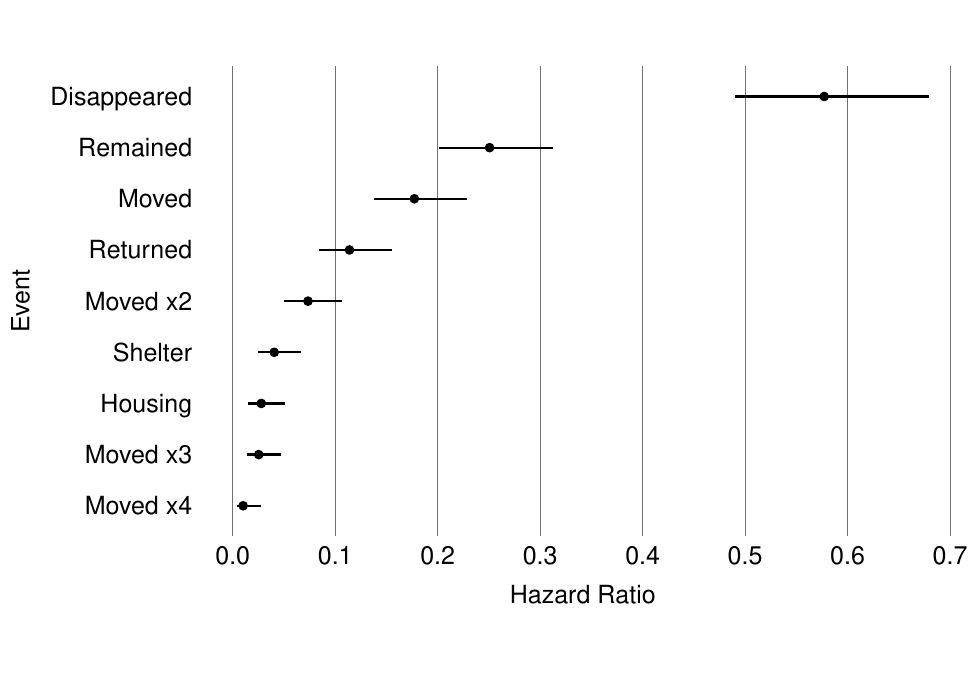}
\caption{\textit{Relative risks of certain events within a year after displacement.} Posterior mode estimates and 95\% credible intervals. Higher ratios suggest greater likelihoods of occurrence.}
\label{fig:rem_full}
\end{figure}

Interestingly, the likelihood of migrating between tracts multiple times seems comparable to moving indoors once (Figure \ref{fig:rem_full}). The risks of traveling to non-origin tracts three (95\% CI 0.014 to 0.047) or four times within a year (0.004 to 0.027) do not appear significantly different from those of entering shelter (0.025 to 0.067) or housing (0.015 to 0.051). Unfortunately, our data cannot corroborate whether clients migrated to these new locations due to additional clearances or voluntary reasons. These results nevertheless still suggest a reality of instability for displaced residents, rather than one of eventual security. Although theoretically illuminating, however, including multiple moves does not improve model fit (BIC 2975 versus 2733). We therefore use the simpler specification for all subsequent analyses.

\begin{figure}[t]
\centering
\includegraphics[scale=0.7]{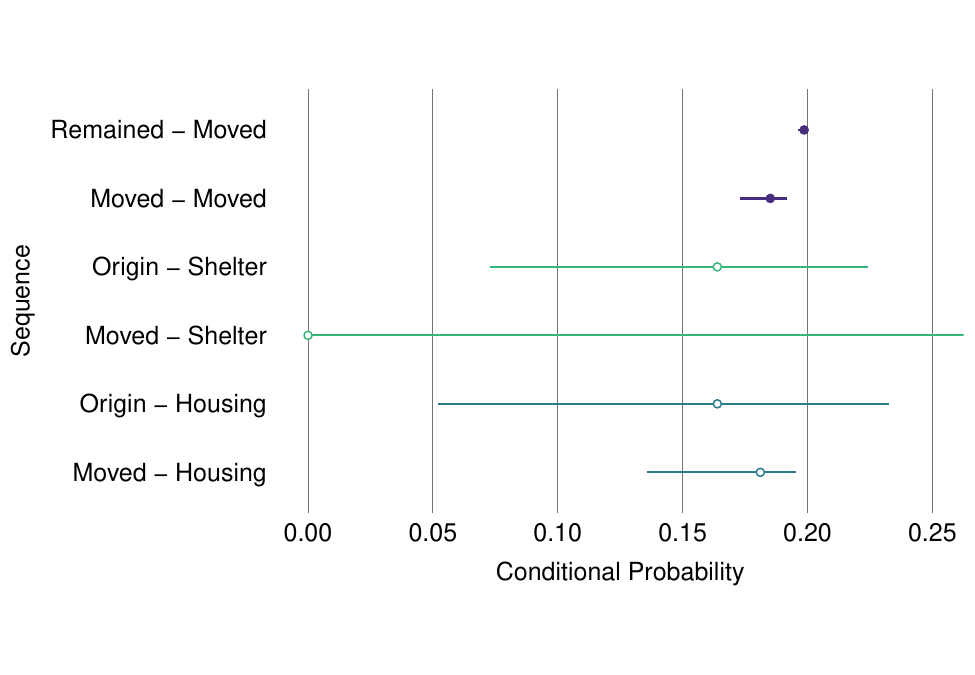}
\vspace{-1em}
\caption{\textit{Conditional probabilities of outcomes given prior events.} Based on posterior modes and 95\% credible intervals. The colors signify directly comparable sequences, with filled circles suggesting significant differences.}
\label{fig:rem_seq}
\end{figure}

Whether a client relocates tracts also impacts the likelihood of certain events. Figure \ref{fig:rem_seq} displays the conditional probabilities of outcomes assuming that a prior event occurred (e.g., entering housing after moving to a non-origin tract). We can directly compare sequences with the same final event (grouped by color). Given that a client remains in their origin tract after eviction, they face a 0.199 probability of moving tracts within the next year (95\% CI 0.196 to 0.200). However, if they already moved tracts, the probability of migrating to another non-origin tract decreases to 0.185 (95\% CI 0.173 to 0.192). Although marginal, this statistically significant difference may suggest that cleared sites remain prone to repeat intervention. Displaced residents may also simply relocate to areas that experience lower enforcement. Nevertheless, whether a client resides in their origin tract does not appear significantly correlated with increased chances of entering shelter or housing. Note that the low overall likelihood of moving indoors contributes to few observations for analysis, which may partly explain the lower precision and higher uncertainty of these effect estimates.

\begin{figure}[t]
\centering
\includegraphics[scale=0.7]{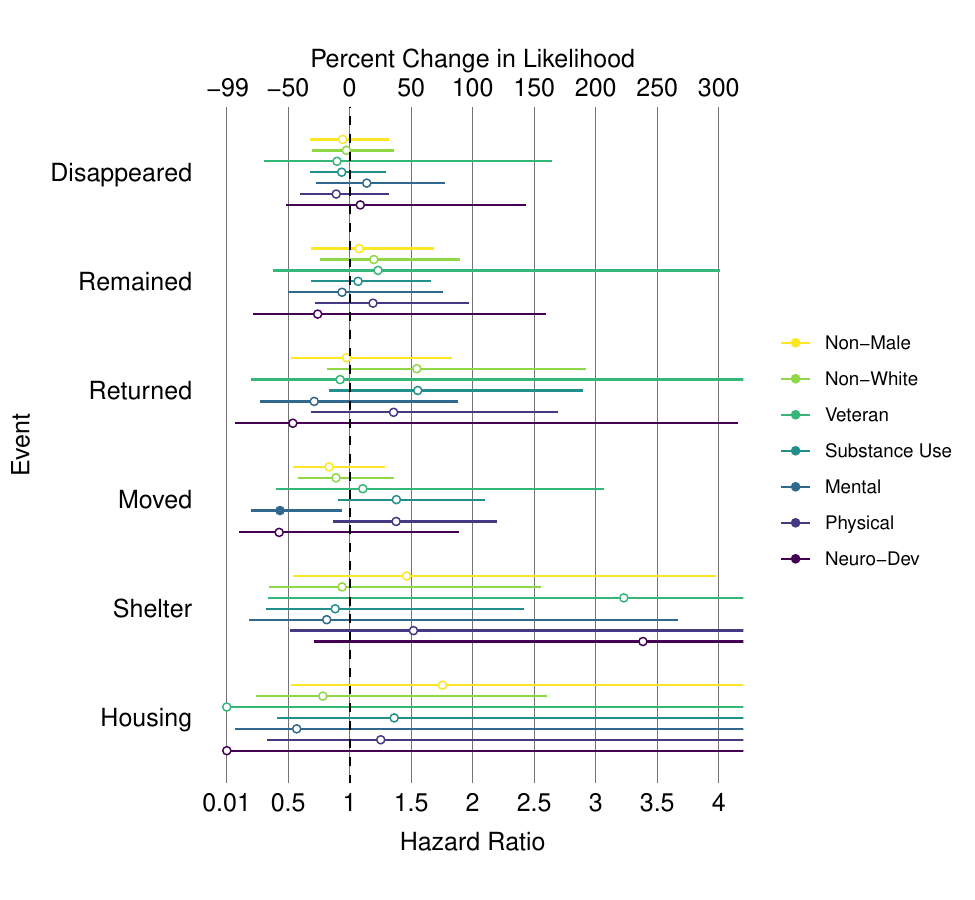}
\caption{\textit{Effects of client characteristics on the likelihoods of outcomes.} The plot shows posterior modes and 95\% credible intervals from seven separate models with one covariate each. Filled circles suggest a statistically significant difference.}
\label{fig:rem_client}
\end{figure}

The likelihoods of outcomes appear relatively consistent across clients, regardless of demographic characteristics (Figure \ref{fig:rem_client}). Although identifying as non-male increases the relative risk of entering shelter by 46.2\% (95\% CI 0.54 to 3.98) and obtaining housing by 75.5\% (95\% CI 0.53 to 5.85) compared to males, these effects do not appear statistically significant. Identifying as non-White similarly does not correlate with different likelihoods. Although non-White clients face an estimated 54\% higher risk of returning to an origin tract compared to White peers, this association appears weak and possibly consistent with a null model of no effect (p = 0.18, 95\% CI 0.82 to 2.92). Veteran status also does not relate to distinct likelihoods of any outcome. We predict a 222.8\% greater risk of entering shelter for veterans compared to non-veterans, but the uncertainty around this estimate remains high (p = 0.31). Again, the small probability of transitioning indoors across all clients could explain the sizable errors in these estimates.\endnote{The veteran model relied on a smaller sample than the baseline specification (167 compared to 395) due to missing data. For demographic covariates, we only included clients for whom information was recorded. As such, the gender and race models also used slightly reduced samples (394 and 378, respectively).} Large but non-significant hazard ratios may indicate low statistical power in these analyses, potentially obscuring the covariates' true effects. Increasing the sample sizes might bring the credible intervals for these attributes into statistically significant ranges.

Individual health seems slightly more predictive of migration patterns (Figure \ref{fig:rem_client}). Most notably, we estimate a statistically significant 56.6\% decrease in the relative risk of moving tracts for clients with mental illness compared to those without (p $<$ 0.05, 95\% CI 0.20 to 0.94). People may not relocate far due to service connections and other place attachments \citep{beckett_banished_2010}, as well as a lack of resources or capacity. We observe similar ties to origin tracts for clients with behavioral health conditions. For example, substance use disorder increases the risk of moving tracts by 38.0\% (95\% CI 0.91 to 2.10) and returning by 55.3\% (95\% CI 0.83 to 2.90). Albeit not significant at an alpha level of 0.05 (p = 0.13 and 0.17, respectively), these weak associations could suggest potential place attachments. Physical illnesses or disability similarly raises the estimated risk of moving tracts by 37.8\% (95\% CI 0.86 to 2.20, p = 0.18) and returning by 35.7\% (95\% CI 0.68 to 2.69, p = 0.38). However, this difference proves less statistically significant compared to that of substance use. Interestingly, having a neurological or developmental condition may not strongly correlate with higher likelihoods of any event. Although we estimate a 238.4\% greater risk for entering shelter (95\% CI 0.71 to 16.2) for those with a neuro-developmental disorder compared to those without one, this association remains weaker (p = 0.13). Shelter and housing outcomes ultimately do not appear linked to individual health based on our existing data.

\begin{figure}[tbp]
\centering
\includegraphics[scale=0.7]{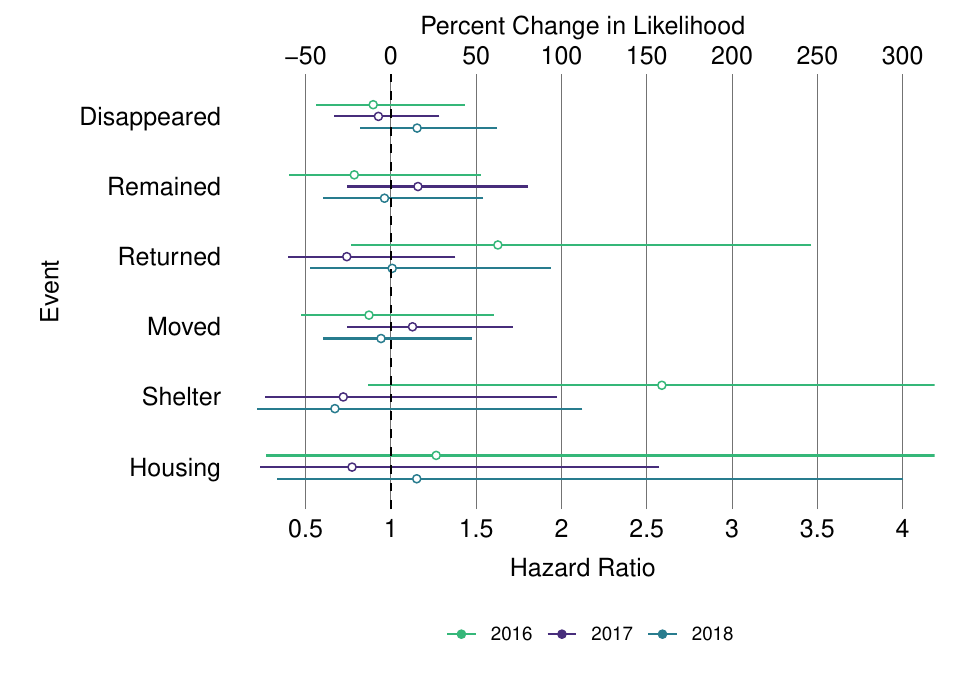}
\caption{\textit{Effects of clearance year on the likelihoods of outcomes.} The plot shows posterior modes and 95\% credible intervals for three separate models with one covariate each. Filled circles suggest a statistically significant difference.}
\label{fig:rem_year}
\end{figure}

The year of removal also does not significantly correlate with different likelihoods for any outcome (Figure \ref{fig:rem_year}). However, we estimate a 158.8\% increase (95\% CI 0.87 to 7.72) in the relative risk of entering shelter for clients displaced in 2016 (i.e., at the Jungle), compared to other periods. Albeit weaker (p $<$ 0.1), this association could reflect the extensive outreach conducted by city and non-profit agencies prior to the Jungle removal. Unlike other sites, which may have only received 72 hours of notice and support, Seattle began planning the Jungle clearance months in advance. Local service organizations likely received more time (even if not additional resources) to engage with unsheltered residents. Furthermore, clients' awareness of the scheduled removal could have prompted preemptive moves for some people. However, among the REACH clients identified at the site up to 90 days prior, only two people moved indoors before the final clearance day -- both at least a week before. In comparison, 6 clients across the rest of the sample (1.5\%) recorded a new indoor residence up to three days before their displacement. This difference appears marginal and suggests that the Jungle's unique situation may not have substantially impacted pathways.

\begin{figure}[tbp]
\centering
\includegraphics[scale=0.7]{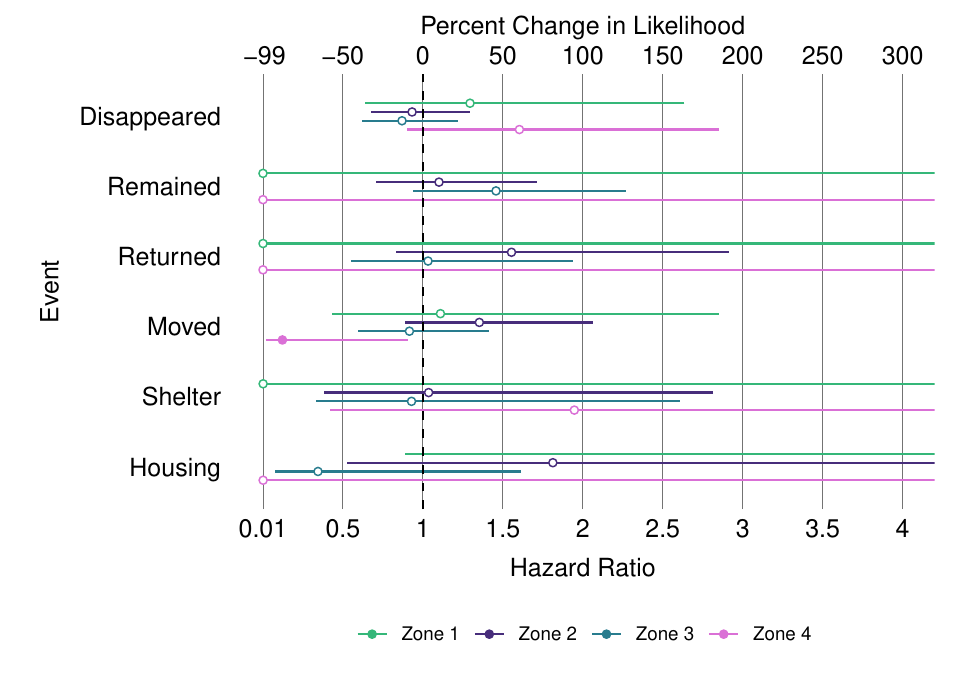}
\caption{\textit{Effects of clearance neighborhood zone on the likelihoods of outcomes.} The plot shows posterior modes and 95\% credible intervals for three separate models with one covariate each. Filled circles suggest a statistically significant difference.}
\label{fig:rem_location}
\end{figure}

Although not strongly predictive of outcomes, displacement location appears weakly associated with migration and housing patterns (Figure \ref{fig:rem_location}). For this analysis, we group neighborhoods into four distinct areas according to geographic proximity and shared characteristics (e.g, land use type, population density). We estimate a 47.5\% increase in the relative risk of remaining in an origin tract for people displaced in Zone 3, compared to elsewhere (95\% CI 0.94 to 2.27, p $<$ 0.1). These removal sites are located around non-residential and industrial areas in SODO (see Figure \ref{fig:ref_map}), which are potentially less exposed to public complaints and law enforcement contact, thus reducing pressures to relocate far. In contrast, Zone 2 correlates with higher likelihoods of moving from (35.3\%, 95\% CI 0.89 to 2.06) and returning to origin tracts (55.4\%, 95\% CI 0.83 to 2.91). Albeit not statistically significant (p = 0.16), the direction of these effects remains notable given that all our clearances concentrated around overpasses and other liminal land. Despite similar spatial attributes, broader contextual factors may still matter. Zone 2 sites are notably closer to densely populated, mixed-use, and gentrifying neighborhoods (e.g., Judkins Park, Beacon Hill, Yesler Terrace). Per complaint- \citep{herring_complaint-oriented_2019} and development-oriented policing models \citep{beck_policing_2020}, these locations could experience disproportionate patrols that encourage repetitive displacement and spatial churn. Lastly, Zones 1 (encompassing Casecade and Eastlake) and 4 (Highland Park and South Park) each consist of a single removal event with fewer clients, resulting in more uncertainty around estimates. Residing in Zone 4 nevertheless appears associated with a 87.8\% decrease in the risk of moving tracts (95\% CI 0.02 to 0.91, p $<$ 0.05). However, this reduction does not imply greater chances of remaining nearby. Clients instead seem to experience a 60.4\% higher likelihood of disappearing (95\% CI 0.90 to 2.85, p = 0.1). Such trends could reflect isolation from social service hubs in central Seattle, as well as proximity to King County and REACH boundaries.

In addition to failing to bring people indoors, clearances do not appear to displace people far. Clients who remain unsheltered and in contact with an HMIS service provider\endnote{For distance analyses, we do not include clients who immediately ``disappeared" from the sample. Clients must record an engagement with REACH or an HMIS service provider within the next year to be counted.} (n = 221) travel an average of 984 meters (m) within 30 days of removal. This represents the cumulative distance between consecutive outdoor encounters with REACH. For example, if staff engage with someone 200 m from their clearance site after a week, then 500 m from that second location on day 30, the individual would travel a total of 700 m. However, most clients migrate a net 0 m between staff encounters, suggesting that they return to and remain near their initial campsite after displacement. Staff may not yet encounter these individuals, given that our analyses assume inertia between engagements. Nonetheless, the median cumulative distance between REACH encounters by 60 days still remains zero. Clients unsheltered after one year (n = 174) have also traveled only a median of 717 m -- less than a ten-minute walk.

\begin{figure}[t]
\centering
\includegraphics[scale=0.8]{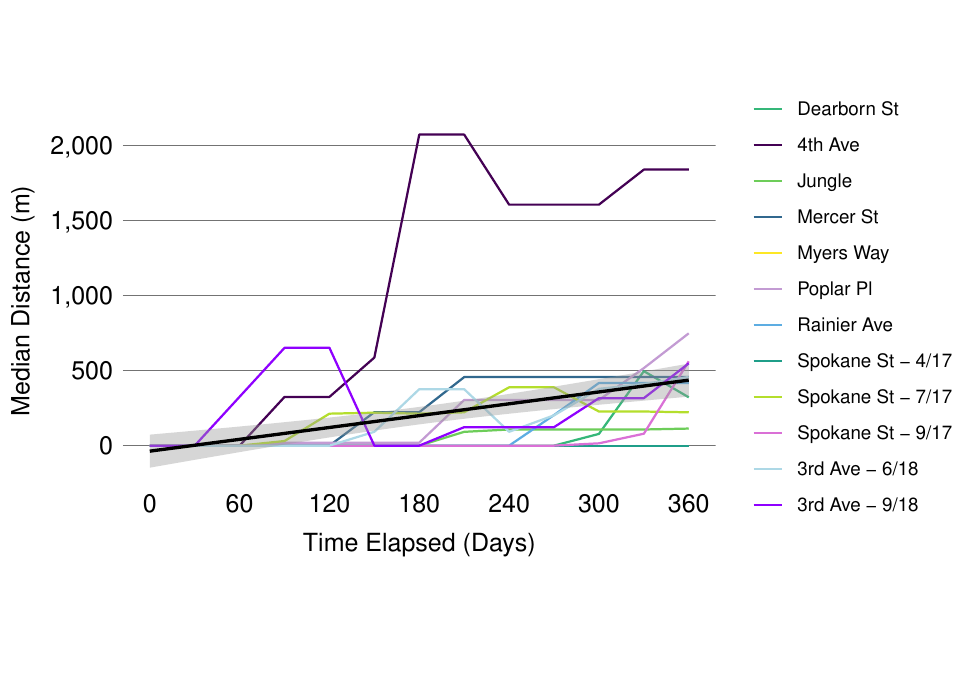}
\vspace{-4em}\caption{\textit{Median distance traveled by clients relative to original location, per removal event.} The black line and ribbon represent a LOESS smoothing curve with a 95\% CI fitted to the group values.}
\label{fig:med_dist_origin}
\end{figure}

These numbers notably reflect total distances migrated regardless of direction. When considering movement relative to initial locations, clients remain even closer. On average, REACH encounters individuals within 960 m of their original site after 30 days and 2.2 km at 60 days. Yet contact with the median client still occurs at the clearance spot (0 m). After one year, at least half of the remaining unsheltered sample engages with outreach staff within 380 m of their displacement sites. These trends furthermore hold across clearance events, with clients staying within a median of 500 m of most removal sites (Figure \ref{fig:med_dist_origin}). For eight clearances, individuals either remain at or return to their initial location within 120 days. Clients' movements thus appear geographically constrained, as documented elsewhere \citep{jocoy_homelessness_2010}. Admittedly, each time sample only includes those who have not yet lost contact with HMIS providers. Clients could theoretically disappear due to remaining highly mobile or leaving the state.

\begin{figure}[t]
\centering
\begin{subfigure}{0.45\textwidth}
\hspace{-2em}\includegraphics[scale=0.5]{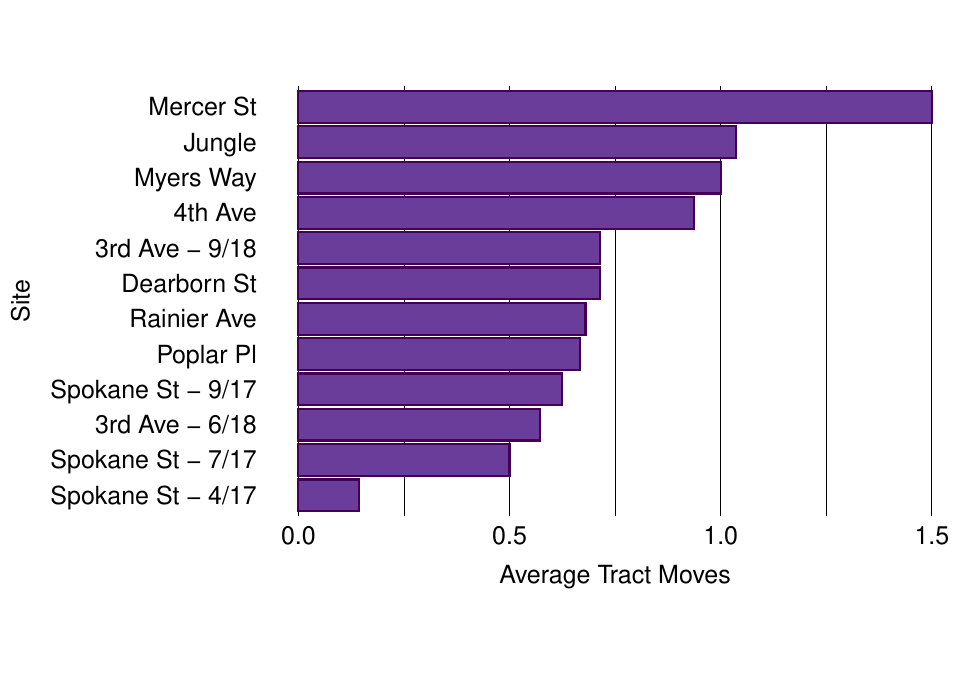}
\end{subfigure}
\begin{subfigure}{0.45\textwidth}
\includegraphics[scale=0.5]{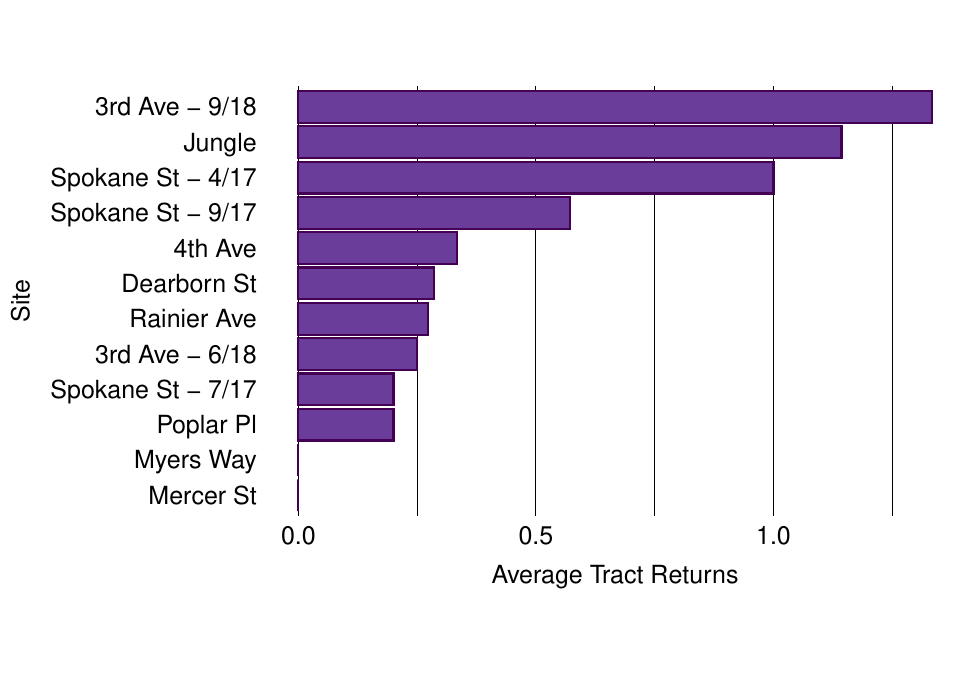}
\end{subfigure}
\vspace{-1em}
\caption{\textit{Average client moves to non-origin tracts (left) and returns to origin tracts (right) within a year, grouped by removal event.}}
\label{fig:client_moves}
\end{figure}

In addition to remaining tethered to small areas, individuals engage with outreach staff in relatively stable coordinate locations. Unsheltered clients who maintain contact with REACH migrate a mean 1.54 times between points in the following year. Yet approximately 32\% of clients do not change coordinates. Furthermore, approximately 14\% of those who move ultimately return to their original site within a year. As anticipated, such relocation may lack geographic significance, given that people rarely cross between Census tracts (an average 0.75 times). Additionally, 53\% of those who leave their initial tract eventually come back within a year post-removal. Although these patterns vary between clearance events, clients from most sites do not typically engage with REACH outside of their origin tract (Figure \ref{fig:client_moves}). Given the REM results, this pattern implies that entering shelter or housing is especially unlikely. Interestingly, individuals from Spokane St and 3rd Ave (located in the industrial neighborhood of SODO) appear least probable to move and more probable to return should they leave. These patterns support our relational event models, which also suggest more spatial stability for these clients -- possibly due to less intensive public and police surveillance in non-residential districts. In contrast, those displaced from Mercer St appear most likely to leave and remain away from their origin tract. This densely populated, mixed-use area near downtown may experience unique enforcement practices, environmental conditions (e.g., less hidden spaces), and frequencies of intervention. Nevertheless, variations could partly reflect unequal tract sizes and site proximity to boundaries.

\begin{figure}[tbp]
\centering
\includegraphics[scale=0.8]{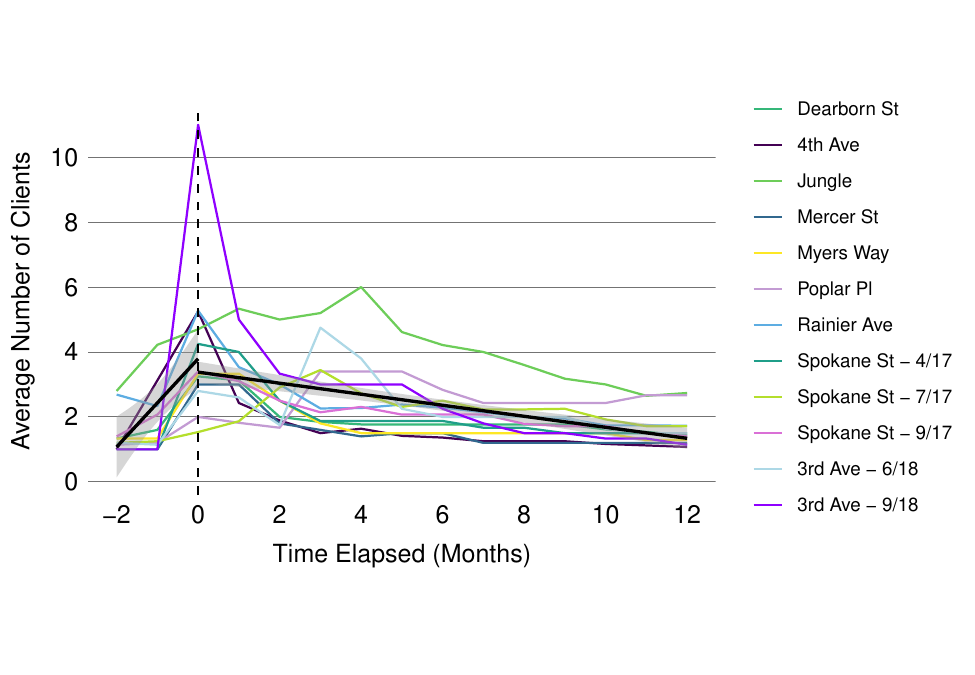}
\vspace{-3em}\caption{\textit{Average number of clients at a point location by shared removal event.}}
\label{fig:client_counts}
\end{figure}

\begin{figure}[tbp]
\centering
\includegraphics[scale=0.8]{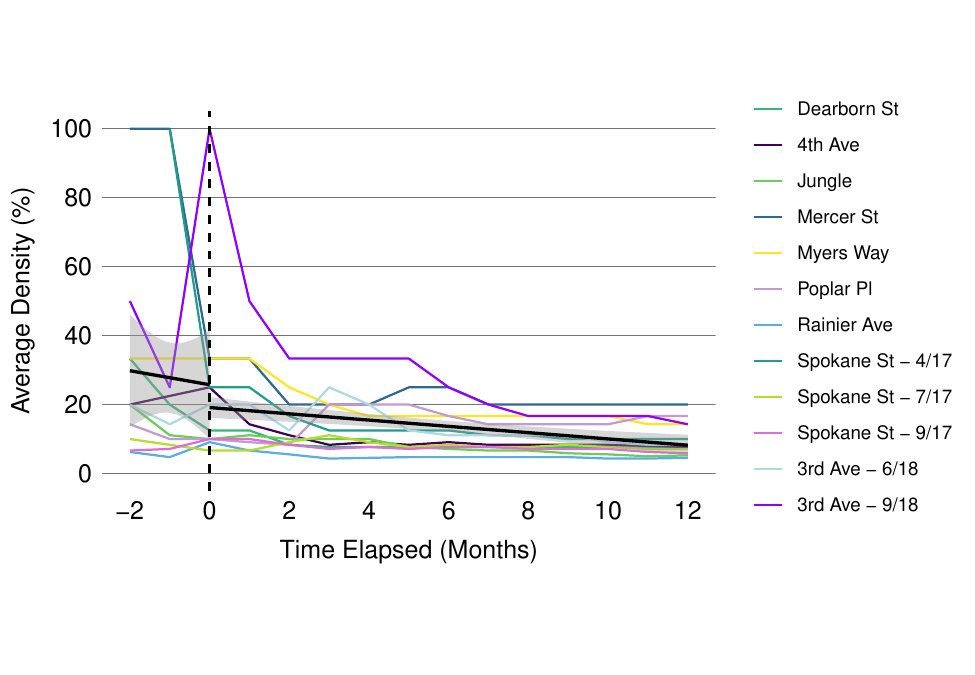}
\vspace{-3em}\caption{\textit{Average population density at a point location by shared removal event.} Values represent the proportion of clients relative to the remaining subsample from their respective clearance.}
\label{fig:client_density}
\end{figure}

\begin{figure}[tbp]
\centering
\includegraphics[scale=0.8]{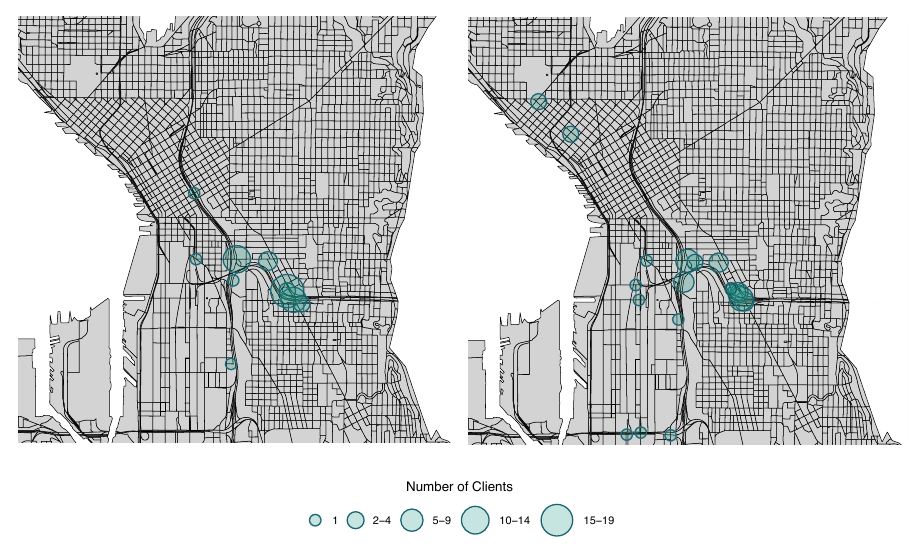}
\caption{\textit{Number of Jungle clients at coordinate locations upon displacement (left) and one year after (right).}}
\label{fig:jungle_map}
\end{figure}

Although not pushing people far, clearances may still sever camp communities and reduce the concentration of homelessness. In the months preceding removal, the average number of residents at a coordinate location steadily increases.\endnote{The average number of clients prior to displacement may not equal the total sample for a removal, given that clearances can affect multiple point locations. The mean density may similarly not cover 100\% of the population.} After displacement, however, clients resettle with fewer people from their original community (Figure \ref{fig:client_counts}). This could result from attrition, as clients disappear or transition indoors. However, we observe similar patterns when comparing numbers of clients to the remaining subsample from a particular removal (referred to here as ``density;" Figure \ref{fig:client_density}). Of course, clients could theoretically move into an encampment where other people already reside. Unfortunately, we cannot test this claim given the lack of representative data on the locations of all people sleeping unsheltered in Seattle. However, our sample still shows a decrease in the mean number and density of clients at coordinate locations over time. Mapping these patterns only further reaffirms the spatial stability, yet disrupted communities, of displaced clients (Figure \ref{fig:jungle_map} for the Jungle and Appendix Figure \ref{figure:combined_maps} for additional sites).

\begin{figure}[t]
\centering
\includegraphics[scale=0.8]{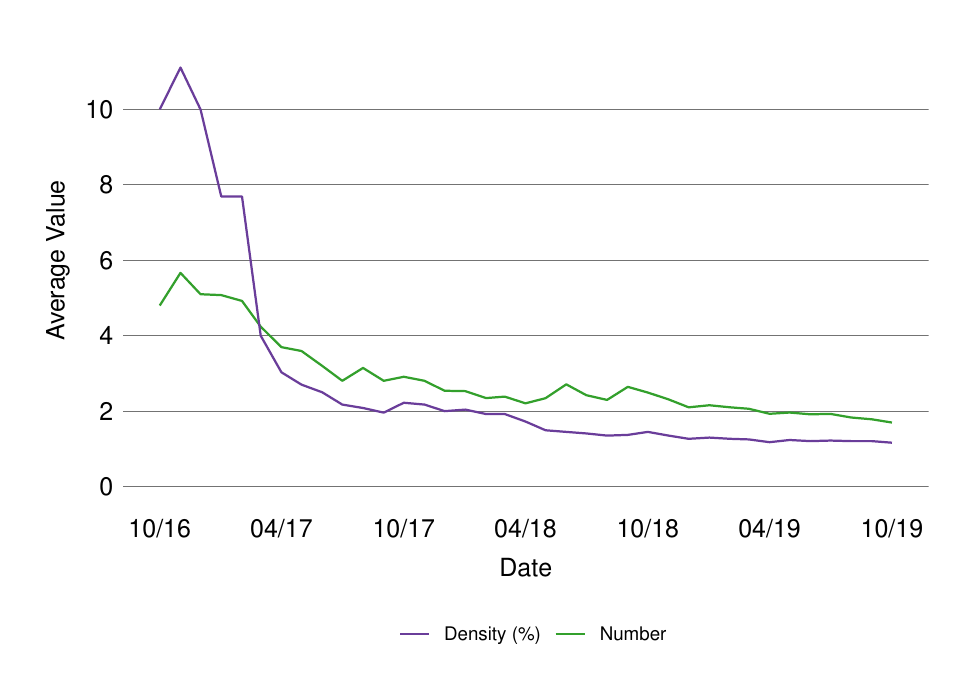}
\caption{\textit{Average number and density of clients by point location.} These values reflect the entire sample from the first removal event through a year after the last one.}
\label{fig:location_density}
\end{figure}

\begin{figure}[t]
\centering
\includegraphics[scale=0.8]{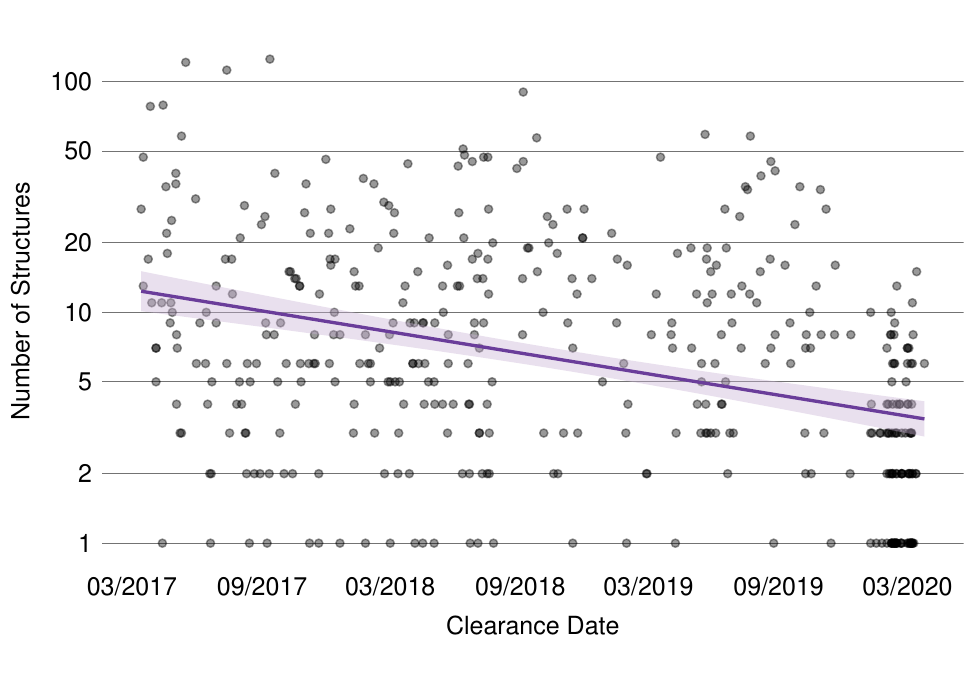}
\vspace{-1em}\caption{\textit{Number of encampment structures at inspection by clearance date.} The purple line and ribbon represent a robust linear trend and 95\% CI.}
\label{fig:clearance_size}
\end{figure}

When considering all unique clients together (instead of by removal event), the size and density of camp populations similarly decline (Figure \ref{fig:location_density}). Although these values occasionally spike due to the addition of new removal events (e.g., in June and September 2018), the overall negative slope suggests that the sample grows increasingly dispersed over time. These trends also reflect patterns documented in publicly available site journals. From 2017 through the start of the COVID-19 pandemic, the typical size of communities displaced by authorities steadily contracts (Figure \ref{fig:clearance_size}). Among clearances in March 2017, officials identified a median of 23 structures at inspection. By March 2020, this value decreased to 2, despite King County's unsheltered population remaining stable across the period. Per Point-In-Time Counts, an estimated 5485 people slept outside on a given night in January 2017, compared to 6320 in 2018, 5228 in 2019, and 5578 in 2020 \citep{allhome}. A street-based census in 2019 and 2020 similarly found that the median Seattle encampment consisted of 3 tents, while approximately one-quarter of locations had a single structure \citep{snedker_tent_2025}. Interestingly, the proportion of sites with larger clusters grew from December 2019 to July 2020 -- coinciding with the COVID-19 pause in clearances. Such trends support our findings that displacement may disrupt the growth of camp populations.

\section*{Discussion}

Although encampment clearances may reduce the visibility of homelessness, displaced residents typically do not move indoors or migrate far. Our findings suggest that clients rarely enter shelter or housing and instead confront high risks of losing contact with service providers. These patterns hold regardless of individual demographics, although clients with mental illness and behavioral health issues demonstrate stronger ties to their original camp areas. Such trends also weakly correlate with displacement location, suggesting that neighborhood conditions could influence these outcomes. Among those who remain unsheltered, people furthermore stay remarkably close to their initial sites, with REACH staff encountering most clients within a ten-minute walk. In fact, the locations of these engagements appear surprisingly stable over time, highlighting the geographically constrained nature of unhoused mobility. Yet clearances may still decrease the concentration of homelessness by severing social networks and reducing camp communities.

With respect to housing outcomes, displaced residents appear most probable to lose service connections and least likely to exit the streets. The high risk of disappearance reaffirms that clearances may succeed in reducing the visibility of homelessness. Unfortunately, however, we cannot determine whether these clients found lodging, relocated to hidden spots within Seattle, or left Washington state. Regardless, the likelihoods of unsheltered events (including shuffling between multiple tracts) are larger than entering shelter or housing. These patterns for transitioning indoors furthermore appear consistent across clients regardless of individual demographics or health conditions. This seems surprising given documented service barriers for those with gender-variant identities, substance use disorder, and disabilities, as well as women and minorities \citep{chang_harms_2022, junejo_no_2016, nichols_supporting_2021, wusinich_if_2019}. Nevertheless, we observe that mental illness correlates with decreased risks of moving tracts, while substance use disorder weakly predicts returning. These clients may exhibit stronger attachments to area resources and less capacity to comply with enforcement \citep{beckett_banished_2010}.

Outcomes may also relate to broader contextual characteristics. Although potentially null, we estimate an increased risk of entering shelter for the 2016 Jungle clearance. Authorities planned this removal for months, and sustained outreach possibly facilitated more shelter referrals. Beginning in 2017, however, the city only required a 72-hour notice and permitted some clearances with less warning, thus posing more obstacles to service connections. Additionally, we demonstrate that location could impact migration patterns. Displacement from Industrial Zone 3 appears positively associated with remaining in place, potentially due to less exposure to public surveillance and police enforcement. In contrast, Zone 2 weakly correlates with moving from and returning to initial tracts. This jurisdiction consists of dense, mixed-use, and gentrifying neighborhoods \citep{hwang_gentrification_2020}. Clients may thus prefer staying here due to accessible amenities \citep{sutton_encampments_2026} -- even if they still sleep in hidden spaces \citep{snedker_tent_2025}. Yet such locations might witness more intervention due to increased complaints \citep{herring_complaint-oriented_2019}, development \citep{beck_policing_2020}, and racial disparities \citep{chen_smartphone_2023}. Together, stronger place attachments and repetitive displacement may encourage more spatial churn between such tracts.

Our results further indicate that displaced residents do not migrate far and typically return to their initial sites, thus corroborating prior interview- and survey-based studies \citep{darrah-okike_it_2018, herring_pervasive_2020}. REACH encounters most unsheltered clients at their original locations within two months of removal. After a year, staff still engage with a third at these same spots, while most others remain within 500 meters. Clients also rarely cross Census tracts, and approximately half of those who leave come back to their initial tracts. Given that Census borders are designed to capture homogeneous population attributes, this spatial inertia could demonstrate individual preferences for certain neighborhoods. It might also reflect physical barriers to movement caused by boundary features (e.g., highways, water bodies, train tracks). Like Herring et al. \citeyearpar{herring_pervasive_2020}, we furthermore do not find evidence of unidirectional movement into certain spaces, but rather a churning between nearby areas. Altogether, these findings highlight the constrained nature of unhoused travel \citep{jocoy_homelessness_2010} and residential mobility \citep{crowder_neighborhood_2011}, as well as people's attachments to area resources \citep{beckett_banished_2010, darrah-okike_it_2018}. More broadly, these patterns reflect other processes of forced displacement, such as eviction -- whereby state action circulates poor tenants between unstable housing within disadvantaged neighborhoods \citep{desmond_evicted_2016}.

Despite not displacing clients far, clearances may still reduce the visible concentration of homelessness. Removal interrupts the growth of camp populations, potentially suggesting that these interventions target more noticeable sites \citep{margier_compassionate_2023}. Furthermore, clearances disrupt social networks and disperse unsheltered residents, as indicated by the decrease in the average number and density of clients at subsequent locations over time. Although people could move to sites with existing occupants, municipal documents similarly demonstrate that individuals may simply settle into smaller groups. Such patterns seem notable given that the estimated unsheltered population in King County remained relatively stable across the study period. This dispersion and isolation also appear concerning because robust encampment communities can offer residents emotional and material support, as well as protect against adverse health issues \citep{boucher_they_2022, chang_harms_2022, junejo_no_2016}. Such ``invisibilization" represents a common theme of US poverty governance. Whether amidst late 19th-century industrialization \citep{riis_how_1890}, post-WWII affluence \citep{harrington_other_1962}, neoliberal economic restructuring \citep{wacquant_punishing_2009}, or global recession \citep{desmond_evicted_2016}, policy choices and structural forces have spatially confined marginalized and minoritized populations away from middle- and upper-class residents.

Several limitations could nonetheless constrain the implications of these findings. Most notably, our study cannot measure the specific effects of clearances on pathways through homelessness, given that our entire sample experienced displacement. Future studies might control for exposure to removal, for example, by comparing unauthorized encampments to sanctioned sites. Secondly, our analysis remains based on one outreach program. Although our data includes engagements with other HMIS-integrated agencies, these providers may not proactively search for people on the streets. Clients could remain unsheltered outside of REACH's catchment area and simply not engage with services. Furthermore, different data-sharing agreements among providers could limit the information available on clients. Scholars must consider these possible system gaps when assessing people's changing resource connections and living situations. Additionally, our analyses focus on a limited subset of clearances -- partly due to model demands for larger camp networks, as well as the labor required to link clients to removal events. By constraining our sample, we lose statistical power for estimating the risks of rarer events (e.g., entering housing). Displacement from smaller encampments post-COVID might also result in different outcomes compared to mass removals under prior contexts. Finally, this paper relies on a single city, and researchers should test whether these patterns hold elsewhere. The localized nature of homelessness, real estate markets, and service geographies could feasibly influence displacement practices.

Nevertheless, our findings still support various policies to prevent displacement and reduce homelessness. At a minimum, we encourage jurisdictions to better track outcomes from interventions. Although Seattle has improved its reporting of shelter referrals from clearance events, data around placement remains lacking. Such information would help determine whether the city's use of outreach services prior to displacement helps bring people indoors or simply serves to justify removal \citep{margier_compassionate_2023, herring_complaint-oriented_2021}. Our study suggests the latter, corroborating critiques that removal may instead serve to manage urban aesthetics. Moving forward, we must continue interrogating the motives behind these tactics, while advocating for those evidenced to increase housing equity. Local governments should ensure affordable housing in neighborhoods where unsheltered residents live, especially given that they often remain near places with historic attachments, support networks, and resource connections \citep{beckett_banished_2010, darrah-okike_it_2018, sutton_encampments_2026}. Authorities should preserve and expand affordable units in these locations (via construction, price stabilization, and renter protections), as well as improve access for people with tenancy barriers \citep{chapple_role_2023, desmond_evicted_2016}. As an interim (albeit inadequate) measure, jurisdictions might establish safe camping and parking areas for clients to more stably connect with services \citep{apha_protecting_2023, herring_new_2014, junejo_no_2016}.

\section*{Conclusion}

Our paper helps clarify the trajectories and experiences of unsheltered residents following encampment clearances. Although literature demonstrates the harmful health and material impacts of these practices, the housing outcomes and migration patterns of those affected remain less clear. Addressing this gap, we innovatively leverage longitudinal street outreach data and relational event models to investigate where individuals move after displacement. The results suggest that people confront low likelihoods of entering shelter or housing and instead face much higher risks of losing contact with service providers. Such findings notably lend little support to official claims that these interventions encourage and facilitate the use of resources.

We additionally observe that people do not relocate far following displacement, echoing broader literature on place attachments and residential mobility in high-poverty settings. Removal instead seems to encourage spatial churn between geographically constrained areas, especially in denser districts with more potential exposure to public surveillance and police contact. Nevertheless, clearances may still reduce the visibility of homelessness by limiting the size of camp communities. Such findings align with long-standing themes in US poverty governance, whereby residential exclusion serves to contain and invisibilize marginalized populations. Altogether, this study ultimately meets the urgent need for a deeper understanding of how clearance strategies impact homelessness and pathways to housing. Forced displacement represents an increasingly common tactic across the country, especially given recent federal and local legislation to criminalize camping in public spaces. Scholars should investigate examples of resistance to these carceral approaches that punish individuals for inequitable structural processes and policy decisions. Exploring such cases could grant valuable insight into strategies for bolstering community solidarity and advancing universal rights to housing.

\clearpage

\theendnotes

\section*{Acknowledgments and Disclosures}

\textbf{Funding}: Partial support for this research came from a Eunice Kennedy Shriver National Institute of Child Health and Human Development research infrastructure grant, P2C HD042828, to the UW Center for Studies in Demography \& Ecology; National Science Foundation CAREER Grant \#SES-2142964; and UW Population Health Initiative Tier 3 Grant. The content is solely the authors' responsibility and does not necessarily represent official NIH or NSF views.

\textbf{Conflicts of interest}: Brandon Morande, Amy Hagopian, and Zack W.\ Almquist have no conflicts of interest to declare. Kim Serry, as an ETS REACH employee, acknowledges the importance of transparency and accountability in scientific research and peer review. ETS REACH employed Serry during the period of analysis. \textbf{Financial Interests}: Serry declares that they have no financial interests, such as stocks, patents, or research grants, that may be perceived as affecting their objectivity in the peer review process. \textbf{Organizational Interests}: ETS REACH has a stake in addressing homelessness in King County and may be impacted by the outcomes of this project.

\textbf{Data availability}: To gain access to the data, interested parties must submit a formal application to ETS REACH for review and approval.

\bibliography{morande,peh}

\clearpage

\appendix

\section*{Appendix}

\setcounter{table}{0}
\setcounter{figure}{0}

\renewcommand{\thetable}{A\arabic{table}}
\renewcommand{\thefigure}{A\arabic{figure}}

\subsection*{Relational Event Model}

This paper models ``egocentric" relational events, which represent sequences of interactions between an actor and their environment \citepappendix{marcum_constructing_2015}, such as moving across space or into housing. Our data includes the temporal order of individuals' actions but not the exact timing between events. We therefore employ an ordinal (rather than interval) model, which estimates the likelihood of an observed event history $A_t$ (occurring by time $t$) as \[
p(A_t \mid \theta, X)
= \prod_{i=1}^{M}
\left[
    \frac{
        \lambda_{a_i A_{\tau(a_{i-1})}{\theta}}
    }{
        \sum_{a' \in \mathbb{A}(A_\tau(a_i))}
        \lambda_{a' A_\tau(a_{i-1})^{\theta}}
    }
\right]^{\epsilon(a_i)} .
\] The probability that action $a_i$ will happen next in an event sequence equals its occurrence rate ($\lambda_{a_i A_{\tau(a_{i-1})}\theta}$) divided by the sum of rates for all possible events (including $a_i$). Following \citeappendix{butts_relational_2008}, we parameterize the rate function $\lambda$ as \[
\lambda(s(a), r(a), c(a), X_a, A_t, \theta)
= exp \left[
\lambda_0 + \theta^T u(s(a), r(a), c(a), X_a, A_t)
\right],
\]where $a$ is the event, $s$ and $r$ are the source and recipient actors (equal here), and $c$ designates the action type. $X$ indicates the set of covariates, $\theta$ the vector of real-valued parameters, and $u$ a p-vector of statistics. We additionally assume that $A_t$ proceeds over time interval $[0, t)$, where 0 represents the onset of risk (i.e., a sweep) for all possible events. We define the clearance as a null event $a_0$ with $\tau(a_0) = 0$, while $M$ signifies the number of non-null events in $A_t$. Although not present in our case, some events could be exogenous if their hazards are unrelated to the observed events \textit{and} not a function of $\theta$. \citeappendix{marcum_constructing_2015} denote endogeneity with $\epsilon$, such that $\epsilon(a_i) = 1$ if $a_i \in A_t$ is endogenous and 0 otherwise. Finally, this equation assumes a piecewise constant model, whereby hazards only change when events occur.

\clearpage

\subsection*{Additional Figures and Tables}
\begin{figure}[!hp]
\centering
\includegraphics[scale=1.5]{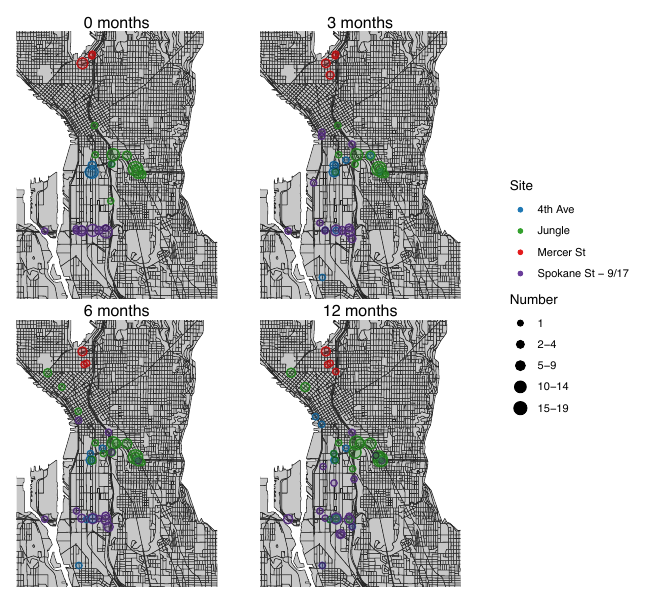}
\caption{\textit{Number of clients at coordinate locations upon and after displacement, grouped by removal event.} The maps show four example sites that do not geographically overlap.}
\label{figure:combined_maps}
\end{figure}

\begin{table}[!hp]
  \caption{Relational Event Model Results - Baseline}
  \label{table:rem_simp}
  \centering
  \begin{threeparttable}
      \begin{tabular}{@{\extracolsep{5pt}}lcccc} 
\\[-1.8ex]\hline 
\hline \\[-1.8ex] 
Outcome & Log Hazard & Std. Dev. & Ratio & 95\% CI \\ 
\hline \\[-1.8ex] 
Displaced (Reference) & 0 & - & 1 & - \\
Disappeared & -0.550*** & 0.083 & 0.577 & (0.490 - 0.679) \\ 
Remained & -1.384*** & 0.112 & 0.251 & (0.201 - 0.312) \\ 
Moved & -1.251*** & 0.107 & 0.286 & (0.232 - 0.353) \\ 
Returned & -2.172*** & 0.157 & 0.114 & (0.084 - 0.155)\\ 
Shelter & -3.206*** & 0.255 & 0.041 & (0.025 - 0.067) \\ 
Housing & -3.581*** & 0.306 & 0.028 & (0.015 - 0.051) \\ 
\hline \\[-1.8ex]
AIC: 2704, BIC: 2733 \\
\hline \\[-1.8ex] 
\multicolumn{5}{l}{\footnotesize\textit{Note:} * $p<.05$, ** $p<.01$, *** $p<.001$, for two-tailed tests. n = 395}
\end{tabular} 
  \end{threeparttable}
\end{table}

\begin{table}[!hp]
  \caption{Relational Event Model Results - Extended}
  \label{table:rem_full}
  \centering
  \begin{threeparttable}
      \begin{tabular}{@{\extracolsep{5pt}}lcccc} 
\\[-1.8ex]\hline 
\hline \\[-1.8ex] 
Outcome & Log Hazard & Std. Dev. & Ratio & 95\% CI \\ 
\hline \\[-1.8ex]
Displaced (Reference) & 0 & - & 1 \\
Disappeared & -0.550*** & 0.083 & 0.577 & (0.490 - 0.679) \\
Remained & -1.384*** & 0.112 & 0.251 & (0.201 - 0.312) \\ 
Moved & -1.730*** & 0.130 & 0.177 & (0.137 - 0.229) \\ 
Moved x2 & -2.612*** & 0.192 & 0.073 & (0.050 - 0.107) \\ 
Moved x3 & -3.676*** & 0.320 & 0.025 & (0.014 - 0.047) \\ 
Moved x4 & -4.592*** & 0.502 & 0.010 & (0.004 - 0.027) \\ 
Returned & -2.172*** & 0.157 & 0.114 & (0.084 - 0.155) \\
Shelter & -3.206*** & 0.255 & 0.041 & (0.025 - 0.067) \\ 
Housing & -3.581*** & 0.306 & 0.028 & (0.015 - 0.051) \\ 
\hline \\[-1.8ex] 
AIC: 2931, BIC: 2975  \\
\hline \\[-1.8ex] 
\multicolumn{5}{l}{\footnotesize\textit{Note:} * $p<.05$, ** $p<.01$, *** $p<.001$, for two-tailed tests. n = 395}
\end{tabular}

  \end{threeparttable}
\end{table}

\begin{table}[!hp]
  \caption{Relational Event Model Results - Sequences}
  \label{table:rem_seq}
  \centering
  \begin{threeparttable}
      \begin{tabular}{@{\extracolsep{5pt}}lccccc} 
\\[-1.8ex]\hline 
\hline \\[-1.8ex] 
Outcome & Log Hazard & Std. Dev. & Ratio & 95\% CI & Cond. Prob.\\ 
\hline \\[-1.8ex]
Displaced (Reference) & 0 & - & 1 & - & - \\
Disappeared & -0.550*** & 0.083 & 0.577 & (0.490 - 0.679) & 0.366  \\ 
Remained & -1.384*** & 0.112 & 0.251 & (0.201 - 0.312) & 0.200  \\ 
Moved & -1.988*** & 0.148 & 0.137 & (0.103 - 0.183) & 0.120  \\ 
Returned & -2.172*** & 0.157 & 0.114 & (0.084 - 0.155) & 0.102  \\ 
Shelter & -3.434*** & 0.293 & 0.032 & (0.018 - 0.057) & 0.031 \\ 
Housing & -4.127*** & 0.411 & 0.016 & (0.007 - 0.036) & 0.016  \\ 
Remained - Moved & 4.649*** & 0.518 & 104.530 & (37.874 - 288.502) & 0.199 \\ 
Moved - Moved & 2.890*** & 0.299 & 17.993 & (10.015 - 32.326) & 0.185 \\ 
Origin - Shelter & 2.542*** & 0.620 & 12.711 & (3.769 - 42.867) & 0.164 \\ 
Moved - Shelter & -6.307 & 33.110 & 0.002 & (0.000 - 2.782e$^{25}$) & 1.875e$^{-5}$ \\ 
Origin - Housing & 2.542** & 0.847 & 12.708 & (2.414 - 66.885) & 0.164\\ 
Moved - Housing & 2.631*** & 0.735 & 13.882 & (3.289 - 58.593) & 0.181 \\ 
\hline \\[-1.8ex] 
AIC: 2509,  BIC: 2567 \\
\hline \\[-1.8ex]
\multicolumn{5}{l}{\footnotesize\textit{Note:} * $p<.05$, ** $p<.01$, *** $p<.001$, for two-tailed tests. n = 395}
\end{tabular}
  \end{threeparttable}
\end{table}

\begin{table}[!hp]
  \caption{Relational Event Model Results - Gender}
  \label{table:rem_gender}
  \centering
  \begin{threeparttable}
      \begin{tabular}{@{\extracolsep{5pt}}lcccc} 
\\[-1.8ex]\hline 
\hline \\[-1.8ex] 
Outcome & Log Hazard & Std. Dev. & Ratio & 95\% CI \\ 
\hline \\[-1.8ex] 
Displaced (Reference) & 0 & - & 1 & - \\
Disappeared & -0.528*** & 0.107 & 0.590 & (0.478 - 0.728) \\ 
Remained & -1.412*** & 0.148 & 0.244 & (0.182 - 0.325) \\ 
Moved & -1.179*** & 0.135 & 0.308 & (0.236 - 0.401) \\ 
Returned & -2.159*** & 0.203 & 0.115 & (0.077 - 0.172) \\
Shelter & -3.376*** & 0.360 & 0.034 & (0.017 - 0.069) \\ 
Housing & -3.846*** & 0.452 & 0.021 & (0.009 - 0.052) \\ 
Disappeared - Non-Male & -0.058 & 0.170 & 0.943 & (0.676 - 1.317) \\ 
Remained - Non-Male & 0.075 & 0.228 & 1.078 & (0.690 - 1.684) \\
Moved - Non-Male & -0.183 & 0.221 & 0.833 & (0.540 - 1.284) \\ 
Returned - Non-Male & -0.025 & 0.321 & 0.975 & (0.520 - 1.830) \\ 
Shelter - Non-Male & 0.380 & 0.510 & 1.462 & (0.538 - 3.977) \\ 
Housing - Non-Male & 0.562 & 0.614 & 1.755 & (0.527 - 5.848) \\ 
\hline \\[-1.8ex] 
AIC: 2709.171, BIC: 2766.867  \\
\hline \\[-1.8ex]
\multicolumn{5}{l}{\footnotesize\textit{Note:} * $p<.05$, ** $p<.01$, *** $p<.001$, for two-tailed tests. n = 394}
\end{tabular}
  \end{threeparttable}
\end{table}

\begin{table}[!hp]
  \caption{Relational Event Model Results - Race}
  \label{table:rem_race}
  \centering
  \begin{threeparttable}
      \begin{tabular}{@{\extracolsep{5pt}} lcccc} 
\\[-1.8ex]\hline 
\hline \\[-1.8ex] 
Outcome & Log Hazard & Std. Dev. & Ratio & 95\% CI \\ 
\hline \\[-1.8ex] 
Displaced (Reference) & 0 & - & 1 & - \\
Disappeared & -0.527*** & 0.121 & 0.590 & (0.465 - 0.749) \\ 
Remained & -1.521*** & 0.175 & 0.219 & (0.155 - 0.308) \\
Moved & -1.166*** & 0.152 & 0.311 & (0.231 - 0.419) \\ 
Returned & -2.376*** & 0.254 & 0.093 & (0.057 - 0.153) \\
Shelter & -3.130*** & 0.361 & 0.044 & (0.022 - 0.089) \\ 
Housing & -3.418*** & 0.415 & 0.033 & (0.015 - 0.074) \\ 
Disappeared - Non-White & -0.027 & 0.170 & 0.973 & (0.698 - 1.357) \\ 
Remained - Non-White & 0.179 & 0.235 & 1.197 & (0.755 - 1.896) \\ 
Moved - Non-White & -0.118 & 0.216 & 0.889 & (0.582 - 1.358) \\ 
Returned - Non-White & 0.435 & 0.324 & 1.546 & (0.819 - 2.918) \\ 
Shelter - Non-White & -0.064 & 0.510 & 0.938 & (0.345 - 2.552) \\  
Housing - Non-White & -0.246 & 0.614 & 0.782 & (0.235 - 2.606) \\ 
\hline \\[-1.8ex] 
AIC: 2618, BIC: 2676 \\
\hline \\[-1.8ex] 
\multicolumn{5}{l}{\footnotesize\textit{Note:} * $p<.05$, ** $p<.01$, *** $p<.001$, for two-tailed tests. n = 378}
\end{tabular}
  \end{threeparttable}
\end{table}

\begin{table}[!hp]
  \caption{Relational Event Model Results - Veteran}
  \label{table:rem_vet}
  \centering
  \begin{threeparttable}
      \begin{tabular}{@{\extracolsep{5pt}} lcccc} 
\\[-1.8ex]\hline 
\hline \\[-1.8ex] 
Outcome & Log Hazard & Std. Dev. & Ratio & 95\% CI \\ 
\hline \\[-1.8ex] 
Displaced (Reference) & 0 & - & 1 & -\\
Disappeared & -0.767*** & 0.143 & 0.465 & (0.351 - 0.614) \\ 
Remained & -1.306*** & 0.174 & 0.271 & (0.193 - 0.381) \\ 
Moved & -0.795*** & 0.144 & 0.452 & (0.341 - 0.599) \\ 
Returned & -1.711*** & 0.205 & 0.181 & (0.121 - 0.270) \\ 
Shelter & -3.657*** & 0.506 & 0.026 & (0.010 - 0.070) \\ 
Housing & -2.646*** & 0.312 & 0.071 & (0.038 - 0.131) \\ 
Disappeared - Veteran & -0.109 & 0.551 & 0.897 & (0.305 - 2.642) \\ 
Remained - Veteran & 0.207 & 0.603 & 1.230 & (0.377 - 4.011)\\ 
Moved - Veteran & 0.102 & 0.520 & 1.107 & (0.399 - 3.070) \\ 
Returned - Veteran & -0.080 & 0.791 & 0.923 & (0.196 - 4.347) \\ 
Shelter - Veteran & 1.172 & 1.157 & 3.228 & (0.334 - 31.204) \\ 
Housing - Veteran & -6.897 & 31.855 & 0.001 & (0.000 - 1.319e$^{24}$) \\  
\hline \\[-1.8ex] 
AIC: 1321, BIC: 1369  \\
\hline \\[-1.8ex]
\multicolumn{5}{l}{\footnotesize\textit{Note:} * $p<.05$, ** $p<.01$, *** $p<.001$, for two-tailed tests. n = 167}
\end{tabular}
  \end{threeparttable}
\end{table}

\begin{table}[!hp]
  \caption{Relational Event Model Results - Substance Use}
  \label{table:rem_sud}
  \centering
  \begin{threeparttable}
      \begin{tabular}{@{\extracolsep{5pt}} lcccc} 
\\[-1.8ex]\hline 
\hline \\[-1.8ex] 
Outcome & Log Hazard & Std. Dev. & Ratio & 95\% CI \\ 
\hline \\[-1.8ex] 
Displaced (Reference) & 0 & - & 1 & - \\
Disappeared & -0.519*** & 0.113 & 0.595 & (0.477 - 0.743) \\ 
Remained & -1.415*** & 0.156 & 0.243 & (0.179 - 0.330) \\ 
Moved & -1.415*** & 0.156 & 0.243 & (0.179 - 0.330) \\ 
Returned & -2.403*** & 0.240 & 0.090 & (0.057 - 0.145) \\ 
Shelter & -3.150*** & 0.340 & 0.043 & (0.022 - 0.084) \\ 
Housing & -3.738*** & 0.452 & 0.024 & (0.010 - 0.058) \\ 
Disappeared - SUD & -0.067 & 0.167 & 0.935 & (0.674 - 1.297) \\ 
Remained - SUD & 0.066 & 0.225 & 1.068 & (0.687 - 1.660) \\ 
Moved - SUD & 0.322 & 0.214 & 1.380 & (0.907 - 2.100) \\ 
Returned - SUD & 0.440 & 0.318 & 1.553 & (0.833 - 2.898) \\ 
Shelter - SUD & -0.125 & 0.514 & 0.883 & (0.322 - 2.417) \\  
Housing - SUD & 0.309 & 0.614 & 1.362 & (0.409 - 4.536) \\ 
\hline \\[-1.8ex] 
AIC: 2711, BIC: 2769  \\
\hline \\[-1.8ex]
\multicolumn{5}{l}{\footnotesize\textit{Note:} * $p<.05$, ** $p<.01$, *** $p<.001$, for two-tailed tests. n = 395}
\end{tabular}
  \end{threeparttable}
\end{table}

\begin{table}[!hp]
  \caption{Relational Event Model Results - Mental Illness}
  \label{table:rem_mental}
  \centering
  \begin{threeparttable}
      \begin{tabular}{@{\extracolsep{5pt}} lcccc} 
\\[-1.8ex]\hline 
\hline \\[-1.8ex] 
Outcome & Log Hazard & Std. Dev. & Ratio & 95\% CI \\ 
\hline \\[-1.8ex] 
Displaced (Reference) & 0 & - & 1 & - \\
Disappeared & -0.570*** & 0.091 & 0.565 & (0.473 - 0.676) \\ 
Remained & -1.374*** & 0.121 & 0.253 & (0.199 - 0.321) \\
Moved & -1.163*** & 0.112 & 0.313 & (0.251 - 0.389) \\ 
Returned & -2.128*** & 0.167 & 0.119 & (0.086 - 0.165) \\ 
Shelter & -3.178*** & 0.273 & 0.042 & (0.024 - 0.071) \\ 
Housing & -3.514*** & 0.321 & 0.030 & (0.016 - 0.056) \\ 
Disappeared - Mental & 0.130 & 0.227 & 1.139 & (0.730 - 1.777) \\ 
Remained - Mental & -0.064 & 0.321 & 0.938 & (0.500 - 1.760) \\
Moved - Mental & -0.835* & 0.393 & 0.434 & (0.201 - 0.937) \\ 
Returned - Mental & -0.340 & 0.495 & 0.712 & (0.270 - 1.878) \\
Shelter - Mental & -0.206 & 0.769 & 0.814 & (0.180 - 3.672) \\ 
Housing - Mental & -0.563 & 1.058 & 0.570 & (0.072 - 4.531) \\ 
\hline \\[-1.8ex] 
AIC: 2709, BIC: 2766  \\
\hline \\[-1.8ex]
\multicolumn{5}{l}{\footnotesize\textit{Note:} * $p<.05$, ** $p<.01$, *** $p<.001$, for two-tailed tests. n = 395}
\end{tabular}
  \end{threeparttable}
\end{table}

\begin{table}[!hp]
  \caption{Relational Event Model Results - Physical Illness}
  \label{table:rem_physical}
  \centering
  \begin{threeparttable}
      \begin{tabular}{@{\extracolsep{5pt}} lcccc} 
\\[-1.8ex]\hline 
\hline \\[-1.8ex] 
Outcome & Log Hazard & Std. Dev. & Ratio & 95\% CI \\ 
\hline \\[-1.8ex] 
Displaced (Reference) & 0 & - & 1 & - \\
Disappeared & -0.524*** & 0.094 & 0.592 & (0.492 - 0.712) \\ 
Remained & -1.427*** & 0.130 & 0.240 & (0.186 - 0.310) \\
Moved & -1.335*** & 0.126 & 0.263 & (0.206 - 0.337) \\ 
Returned & -2.251*** & 0.186 & 0.105 & (0.073 - 0.152) \\ 
Shelter & -3.319*** & 0.307 & 0.036 & (0.020 - 0.066) \\ 
Housing & -3.638*** & 0.358 & 0.026 & (0.013 - 0.053) \\ 
Disappeared - Physical & -0.116 & 0.202 & 0.891 & (0.600 - 1.323) \\ 
Remained - Physical & 0.174 & 0.258 & 1.190 & (0.718 - 1.972) \\ 
Moved - Physical & 0.321 & 0.239 & 1.378 & 	(0.863 - 2.201) \\ 
Returned - Physical & 0.305 & 0.350 & 1.357 & (0.684 - 2.695) \\ 
Shelter - Physical & 0.418 & 0.552 & 1.518 & (0.514 - 4.483) \\ 
Housing - Physical & 0.225 & 0.687 & 1.253 & (0.326 - 4.819) \\ 
\hline \\[-1.8ex] 
AIC: 2712, BIC: 2770  \\
\hline \\[-1.8ex]
\multicolumn{5}{l}{\footnotesize\textit{Note:} * $p<.05$, ** $p<.01$, *** $p<.001$, for two-tailed tests. n = 395}
\end{tabular}
  \end{threeparttable}
\end{table}

\begin{table}[!hp]
  \caption{Relational Event Model Results - Neuro-Developmental}
  \label{table:rem_neuro}
  \centering
  \begin{threeparttable}
      \begin{tabular}{@{\extracolsep{5pt}} lcccc} 
\\[-1.8ex]\hline 
\hline \\[-1.8ex] 
Outcome & Log Hazard & Std. Dev. & Ratio & 95\% CI \\ 
\hline \\[-1.8ex] 
Displaced (Reference) & 0 & - & 1 & - \\
Disappeared & -0.553*** & 0.085 & 0.575 & (0.487 - 0.679) \\ 
Remained & -1.373*** & 0.114 & 0.253 & (0.202 - 0.317) \\ 
Moved & -1.228*** & 0.108 & 0.293 & (0.237 - 0.362) \\ 
Returned & -2.153*** & 0.159 & 0.116 & (0.085 - 0.159) \\ 
Shelter & -3.298*** & 0.272 & 0.037 & (0.022 - 0.063) \\ 
Housing & -3.540*** & 0.306 & 0.029 & (0.016 - 0.053) \\ 
Disappeared - Neuro-Dev & 0.083 & 0.412 & 1.087 & (0.485 - 2.436) \\ 
Remained - Neuro-Dev & -0.301 & 0.639 & 0.740 & (0.211 - 2.592) \\
Moved - Neuro-Dev & -0.851 & 0.758 & 0.427 & (0.097 - 1.885) \\ 
Returned - Neuro-Dev & -0.619 & 1.043 & 0.538 & (0.070 - 4.158) \\ 
Shelter - Neuro-Dev & 1.219 & 0.798 & 3.384 & (0.708 - 16.163) \\ 
Housing - Neuro-Dev & -6.370 & 32.971 & 0.002 & (1.470e$^{-31}$ - 1.990e$^{25}$)\\ 
\hline \\[-1.8ex] 
AIC: 2711, BIC: 2769 \\
\hline \\[-1.8ex]
\multicolumn{5}{l}{\footnotesize\textit{Note:} * $p<.05$, ** $p<.01$, *** $p<.001$, for two-tailed tests. n = 395}
\end{tabular}
  \end{threeparttable}
\end{table}

\begin{table}[!hp]
  \caption{Relational Event Model Results - 2016}
  \label{table:rem_2016}
  \centering
  \begin{threeparttable}
      \begin{tabular}{@{\extracolsep{5pt}} lcccc} 
\\[-1.8ex]\hline 
\hline \\[-1.8ex] 
Outcome & Log Hazard & Std. Dev. & Ratio & 95\% CI \\ 
\hline \\[-1.8ex] 
Displaced (Reference) & 0 & - & 1 & - \\
Disappeared & -0.534*** & 0.090 & 0.586 & (0.492 - 0.699) \\ 
Remained & -1.351*** & 0.120 & 0.259 & (0.205 - 0.328) \\ 
Moved & -1.232*** & 0.115 & 0.292 & (0.233 - 0.365) \\ 
Returned & -2.262*** & 0.178 & 0.104 & (0.074 - 0.148) \\ 
Shelter & -3.419*** & 0.306 & 0.033 & 	(0.018 - 0.060) \\ 
Housing & -3.620*** & 0.338 & 0.027 & (0.014 - 0.052) \\ 
Disappeared - 2016 & -0.110 & 0.239 & 0.896 (0.561 - 1.432) \\ 
Remained - 2016 & -0.241 & 0.339 & 0.786 & (0.404 - 1.526) \\ 
Moved - 2016 & -0.137 & 0.311 & 0.872 & (0.474 - 1.604) \\ 
Returned - 2016 & 0.487 & 0.385 & 1.627 & (0.765 - 3.463) \\ 
Shelter - 2016 & 0.951 & 0.558 & 2.588 & (0.868 - 7.720) \\  
Housing - 2016 & 0.235 & 0.794 & 1.265 & (0.267 - 6.003) \\ 
\hline \\[-1.8ex] 
AIC: 2711, BIC: 2768  \\
\hline \\[-1.8ex]
\multicolumn{5}{l}{\footnotesize\textit{Note:} * $p<.05$, ** $p<.01$, *** $p<.001$, for two-tailed tests. n = 395}
\end{tabular}
  \end{threeparttable}
\end{table}

\begin{table}[!hp]
  \caption{Relational Event Model Results - 2017}
  \label{table:rem_2017}
  \centering
  \begin{threeparttable}
      \begin{tabular}{@{\extracolsep{5pt}} lcccc} 
\\[-1.8ex]\hline 
\hline \\[-1.8ex] 
Outcome & Log Hazard & Std. Dev. & Ratio & 95\% CI \\ 
\hline \\[-1.8ex] 
Displaced (Reference) & 0 & - & 1 & - \\
Disappeared & -0.511*** & 0.118 & 0.600 & (0.476 - 0.757) \\ 
Remained & -1.463*** & 0.167 & 0.232 & (0.167 - 0.321) \\ 
Moved & -1.315*** & 0.158 & 0.268 & (0.197 - 0.366) \\ 
Returned & -2.028*** & 0.213 & 0.132 & (0.087 - 0.200) \\ 
Shelter & -3.050*** & 0.341 & 0.047 & (0.024 - 0.092) \\ 
Housing & -3.455*** & 0.415 & 0.032 & (0.014 - 0.071) \\
Disappeared - 2017 & -0.076 & 0.166 & 0.927 & (0.669 - 1.284) \\ 
Remained - 2017 & 0.147 & 0.226 & 1.159 & (0.744 - 1.804) \\
Moved - 2017 & 0.119 & 0.214 & 1.127 & (0.740 - 1.715) \\
Returned - 2017 & -0.299 & 0.316 & 0.741 & (0.399 - 1.379) \\ 
Shelter - 2017 & -0.327 & 0.514 & 0.721 & (0.263 - 1.974) \\ 
Housing - 2017 & -0.258 & 0.614 & 0.772 & (0.232 - 2.572) \\ 
\hline \\[-1.8ex] 
AIC: 2713, BIC: 2771 \\
\hline \\[-1.8ex]
\multicolumn{5}{l}{\footnotesize\textit{Note:} * $p<.05$, ** $p<.01$, *** $p<.001$, for two-tailed tests. n = 395}
\end{tabular}
  \end{threeparttable}
\end{table}

\begin{table}[!hp]
  \caption{Relational Event Model Results - 2018}
  \label{table:rem_2018}
  \centering
  \begin{threeparttable}
      \begin{tabular}{@{\extracolsep{5pt}} lcccc} 
\\[-1.8ex]\hline 
\hline \\[-1.8ex] 
Outcome & Log Hazard & Std. Dev. & Ratio & 95\% CI \\ 
\hline \\[-1.8ex] 
Displaced (Reference) & 0 & - & 1 & - \\
Disappeared & -0.599*** & 0.103 & 0.549 & (0.449 - 0.673) \\ 
Remained & -1.371*** & 0.137 & 0.254 & (0.194 - 0.332) \\ 
Moved & -1.232*** & 0.130 & 0.292 & (0.226 - 0.376) \\ 
Returned & -2.175*** & 0.193 & 0.114 & (0.078 - 0.166) \\
Shelter & -3.091*** & 0.295 & 0.045 & (0.025 - 0.081) \\ 
Housing & -3.630*** & 0.383 & 0.027 & (0.013 - 0.056) \\ 
Disappeared - 2018 & 0.143 & 0.174 & 1.154 & (0.820 - 1.623) \\ 
Remained - 2018 & -0.038 & 0.240 & 0.963 & (0.601 - 1.541) \\ 
Moved - 2018 & -0.060 & 0.228 & 0.942 & (0.602 - 1.474) \\ 
Returned - 2018 & 0.008 & 0.334 & 1.008 & (0.524 - 1.938) \\ 
Shelter - 2018 & -0.398 & 0.587 & 0.672 & (0.213 - 2.123) \\  
Housing - 2018 & 0.141 & 0.636 & 1.151 & (0.331 - 4.004) \\ 
\hline \\[-1.8ex] 
AIC: 2714, BIC: 2772 \\
\hline \\[-1.8ex] 
\multicolumn{5}{l}{\footnotesize\textit{Note:} * $p<.05$, ** $p<.01$, *** $p<.001$, for two-tailed tests. n = 395}
\end{tabular}
  \end{threeparttable}
\end{table}

\begin{table}[!hp]
  \caption{Relational Event Model Results - Neighborhood Zone 1}
  \label{table:rem_zone1}
  \centering
  \begin{threeparttable}
      \begin{tabular}{@{\extracolsep{5pt}} lcccc} 
\\[-1.8ex]\hline 
\hline \\[-1.8ex] 
Outcome & Log Hazard & Std. Dev. & Ratio & 95\% CI \\ 
\hline \\[-1.8ex] 
Displaced (Reference) & 0 & - & 1 & - \\
Disappeared & -0.564*** & 0.086 & 0.569 & (0.481 - 0.673) \\  
Remained & -1.334*** & 0.113 & 0.263 & (0.211 - 0.329) \\ 
Moved & -1.257*** & 0.110 & 0.285 & (0.230 - 0.353) \\ 
Returned & -2.123*** & 0.158 & 0.120 & (0.088 - 0.163) \\ 
Shelter & -3.157*** & 0.255 & 0.043 & (0.026 - 0.070) \\ 
Housing & -3.732*** & 0.337 & 0.024 & (0.012 - 0.046) \\ 
Disappeared - Zone 1 & 0.258 & 0.362 & 1.295 & (0.636 - 2.635) \\ 
Remained - Zone 1 & -8.463 & 29.108 & 0.0002 & (0 - 1.263$e^{21}$) \\ 
Moved - Zone 1 & 0.104 & 0.481 & 1.110 & (0.432 - 2.849) \\ 
Returned - Zone 1 & -7.761 & 30.247 & 0.0004 & (0 - 2.379$e^{22}$) \\ 
Shelter - Zone 1 & -6.851 & 31.946 & 0.001 & (0 - 1.650$e^{24}$) \\ 
Housing - Zone 1 & 1.481 & 0.816 & 4.398 & (0.888 - 21.781) \\  % p = 0.070
\hline \\[-1.8ex] 
AIC: 2697, BIC: 2755 \\
\hline \\[-1.8ex]
\multicolumn{5}{l}{\footnotesize\textit{Note:} * $p<.05$, ** $p<.01$, *** $p<.001$, for two-tailed tests. n = 395}
\end{tabular}
  \end{threeparttable}
\end{table}

\begin{table}[!hp]
  \caption{Relational Event Model Results - Neighborhood Zone 2}
  \label{table:rem_zone2}
  \centering
  \begin{threeparttable}
      \begin{tabular}{@{\extracolsep{5pt}} lcccc} 
\\[-1.8ex]\hline 
\hline \\[-1.8ex] 
Outcome & Log Hazard & Std. Dev. & Ratio & 95\% CI \\ 
\hline \\[-1.8ex] 
Displaced (Reference) & 0 & - & 1 & - \\
Disappeared & -0.516*** & 0.115 & 0.597 & (0.476 - 0.748) \\ 
Remained & -1.432*** & 0.161 & 0.239 & (0.174 - 0.327) \\ 
Moved & -1.411*** & 0.159 & 0.244 & (0.178 - 0.333) \\ 
Returned & -2.413*** & 0.246 & 0.090 & (0.055 - 0.145) \\ 
Shelter & -3.224*** & 0.361 & 0.040 & (0.020 - 0.081) \\ 
Housing & -3.917*** & 0.505 & 0.020 & (0.007 - 0.054) \\ 
Disappeared - Zone 2 & -0.070 & 0.166 & 0.932 & (0.673 - 1.292) \\ 
Remained - Zone 2 & 0.096 & 0.225 & 1.101 & (0.708 - 1.711) \\ 
Moved - Zone 2 & 0.302 & 0.215 & 1.353 & (0.888 - 2.062) \\  % p = 0.16
Returned - Zone 2 & 0.441 & 0.320 & 1.554 & (0.829 - 2.913) \\ % p = 0.16
Shelter - Zone 2 & 0.035 & 0.510 & 1.036 & (0.381 - 2.815) \\ 
Housing - Zone 2 & 0.595 & 0.635 & 1.813 & (0.522 - 6.290) \\ 
\hline \\[-1.8ex] 
AIC: 2711, BIC: 2769 \\
\hline \\[-1.8ex]
\multicolumn{5}{l}{\footnotesize\textit{Note:} * $p<.05$, ** $p<.01$, *** $p<.001$, for two-tailed tests. n = 395}
\end{tabular}
  \end{threeparttable}
\end{table}

\begin{table}[!hp]
  \caption{Relational Event Model Results - Neighborhood Zone 3}
  \label{table:rem_zone3}
  \centering
  \begin{threeparttable}
      \begin{tabular}{@{\extracolsep{5pt}} lcccc} 
\\[-1.8ex]\hline 
\hline \\[-1.8ex] 
Outcome & Log Hazard & Std. Dev. & Ratio & 95\% CI \\ 
\hline \\[-1.8ex] 
Displaced (Reference) & 0 & - & 1 & - \\
Disappeared & -0.497*** & 0.105 & 0.608 & (0.495 - 0.747) \\ 
Remained & -1.549*** & 0.154 & 0.213 & (0.157 - 0.287) \\ 
Moved & -1.218*** & 0.135 & 0.296 & (0.227 - 0.386) \\ 
Returned & -2.185*** & 0.203 & 0.113 & (0.076 - 0.167) \\ 
Shelter & -3.178*** & 0.323 & 0.042 & (0.022 - 0.078) \\ 
Housing & -3.283*** & 0.340 & 0.038 & (0.019 - 0.073) \\ 
Disappeared - Zone 3 & -0.140 & 0.172 & 0.870 & (0.620 - 1.219) \\ 
Remained - Zone 3 & 0.377 & 0.226 & 1.457 & (0.936 - 2.269) \\  % p = 0.096
Moved - Zone 3 & -0.088 & 0.220 & 0.916 & (0.595 - 1.410) \\ 
Returned - Zone 3 & 0.032 & 0.321 & 1.032 & (0.550 - 1.938) \\ 
Shelter - Zone 3 & -0.074 & 0.527 & 0.929 & (0.331 - 2.608) \\ 
Housing - Zone 3 & -1.067 & 0.788 & 0.344 & (0.073 - 1.614) \\  % p = 0.18
\hline \\[-1.8ex] 
AIC: 2709, BIC: 2767 \\
\hline \\[-1.8ex]
\multicolumn{5}{l}{\footnotesize\textit{Note:} * $p<.05$, ** $p<.01$, *** $p<.001$, for two-tailed tests. n = 395}
\end{tabular}

  \end{threeparttable}
\end{table}

\begin{table}[!hp]
  \caption{Relational Event Model Results - Neighborhood Zone 4}
  \label{table:rem_zone4}
  \centering
  \begin{threeparttable}
      \begin{tabular}{@{\extracolsep{5pt}} lcccc} 
\\[-1.8ex]\hline 
\hline \\[-1.8ex] 
Outcome & Log Hazard & Std. Dev. & Ratio & 95\% CI \\ 
\hline \\[-1.8ex] 
Displaced (Reference) & 0 & - & 1 & - \\
Disappeared & -0.590*** & 0.087 & 0.554 & (0.467 - 0.658) \\  
Remained & -1.313*** & 0.113 & 0.269 & (0.215 - 0.336) \\
Moved & -1.190*** & 0.108 & 0.304 & (0.246 - 0.376) \\ 
Returned & -2.101*** & 0.158 & 0.122 & (0.090 - 0.167) \\ 
Shelter & -3.269*** & 0.272 & 0.038 & (0.022 - 0.065) \\ 
Housing & -3.510*** & 0.306 & 0.030 & (0.016 - 0.054) \\ 
Disappeared - Zone 4 & 0.472 & 0.294 & 1.604 & (0.902 - 2.852) \\  % p = 0.11
Remained - Zone 4 & -8.797 & 28.609 & 0.0002 & (0 - 3.404$e^{20}$) \\ 
Moved - Zone 4 & -2.106* & 1.024 & 0.122 & (0.016 - 0.906) \\ 
Returned - Zone 4 & -8.092 & 29.693 & 0.0003 & (0 - 5.766$e^{21}$) \\ 
Shelter - Zone 4 & 0.666 & 0.782 & 1.947 & (0.421 - 9.013) \\ 
Housing - Zone 4 & -6.850 & 31.949 & 0.001 & (0 - 1.663$e^{24}$) \\ 
\hline \\[-1.8ex] 
AIC: 2681, BIC: 2738 \\
\hline \\[-1.8ex]
\multicolumn{5}{l}{\footnotesize\textit{Note:} * $p<.05$, ** $p<.01$, *** $p<.001$, for two-tailed tests. n = 395}
\end{tabular}

  \end{threeparttable}
\end{table}

\clearpage

\bibliographyappendix{morande,peh}

\end{document}